\documentclass[reprint,amsmath,amssymb,aps,nolongbibliography,nofootinbib,superscriptaddress]{revtex4-2}
\pdfoutput=1
\usepackage{mathtools}
\usepackage{graphicx}% Include figure files
\usepackage{subfig}
\usepackage{dcolumn}% Align table columns on decimal point
\usepackage{bm}% bold math
\usepackage{hyperref}% add hypertext capabilities
\usepackage{xcolor}
\usepackage{appendix}
\usepackage[sfdefault=cmbr]{isomath}
\usepackage{comment}
\usepackage{subcaption}
\usepackage{ragged2e}
\DeclareCaptionJustification{justified}{\justifying}
\setcitestyle{super}
\begin{document}

\title{Shear Layers and Plugs in the Capillary Flow of Wormlike Micellar Gels}

\author{Ronak Gupta}
\affiliation{Department of Mechanical Engineering, University of British Columbia, 6250 Applied Science Ln, Vancouver, British Columbia V6T1Z4, Canada}
\author{Masoud Daneshi}
\affiliation{Department of Mathematics, University of British Columbia, Vancouver, BC V6T 1Z2, Canada}
\author{Ian Frigaard}
\affiliation{Department of Mechanical Engineering, University of British Columbia, 6250 Applied Science Ln, Vancouver, British Columbia V6T1Z4, Canada}
\affiliation{Department of Mathematics, University of British Columbia, Vancouver, BC V6T 1Z2, Canada}
\author{Gwynn J. Elfring} \email{gelfring@mech.ubc.ca}
\affiliation{Department of Mechanical Engineering, University of British Columbia, 6250 Applied Science Ln, Vancouver, British Columbia V6T1Z4, Canada}

\date{\today}

\begin{abstract}
Wormlike micellar solutions formed by long-chained zwitterionic surfactants show gel-like rheology at room temperature and have recently been found to exhibit other complex and interesting rheological features. We study the dynamics of these wormlike micellar gels in a pipe-flow scenario using optical coherence tomography-based velocimetry and report the existence of plug flows with strong wall slip and non-parabolic velocity profiles for different surfactant concentrations and imposed flow rates.  We rationalize these results as features of a developing transient flow of a viscoelastic solution in space and time and show that these shear layers indicate a flow induced heterogeneity. Our experiments shed light on the transient fluid dynamics of wormlike micelles in simple geometries and highlight the complexity of flows involving wormlike micellar gels and similar soft matter systems in canonical flows.
\end{abstract}

\maketitle

\section{Introduction}

Surfactants in solvents tend to aggregate to form a variety of structures owing to their dual nature, namely a hydrophilic head section and hydrophobic tail section.\cite{nagarajan1991theory} These take the shape of spheres, rods and more topologically complex structures that can be tuned by a variety of physio-chemical properties.\cite{dreiss2017wormlike} Under certain favourable conditions, surfactants aggregate to form linear chains - flexible micelles that can then entangle to form a network of \textit{wormlike} micelles (WLM).  This network lends the suspension a microstructure at the mesoscopic level, which influences its rheology and mechanical behaviour.\cite{fardin2014flows}
 
Owing to comparatively weaker bonds when compared with polymeric systems, WLM solutions allow the exchange of monomers at equilibrium enabling micelles to continuously break and recombine. This provides an additional mode of relaxation at equilibrium.\cite{cates1987reptation} If the breaking time is orders of magnitude smaller than the reptation time, a WLM solution behaves like a classical Maxwell viscoelastic fluid where a \textit{single} relaxation time demarcates its solid-like and liquid-like behaviour \cite{cates1987reptation}. This feature makes WLM solutions unique in the broad class of complex fluids and has encouraged a large variety of investigations into their linear and nonlinear rheological behaviour. \cite{rehage1991viscoelastic,spenley1993nonlinear,berret1993linear,berret1997transient,sood1999linear,cates2006rheology}

The combination of viscoelasticity and dynamic microstructure makes WLM solutions a fertile system to explore many interesting instabilities and dynamical behaviour.\cite{fardin2012instabilities,fardin2012interplay,fardin2014flows,perge2014surfactant,rothstein2020complex} For instance, prototypical flows like flow over a sphere,\cite{chen2004flow,sasmal2021unsteady} through a bend\cite{hwang2017flow} and in a cross-slot,\cite{haward2012extensional,dubash2012elastic,kalb2017role} all display significant deviations from their Newtonian counterparts. In a simple-shear flow scenario, semi-dilute WLM solutions can show a nonmonotonic flow-curve. This is an indication of a shear-banding instability in the system,\cite{spenley1996nonmonotonic,britton1997two,olmsted2008perspectives,divoux2016shear}as a result of which the flow becomes inhomogeneous and separates into distinct regions with different properties and shear-rates. Multiple studies have used a cylindrical Couette geometry (to generate simple shear) to illuminate the complex spatio-temporal dynamics involved in shear-banding both at a transient level \cite{lerouge1998shear,becu2004spatiotemporal,decruppe2006local,becu2007evidence,miller2007transient} and in steady state.\cite{salmon2003shear,hu2005kinetics} Finally, WLM solutions in turbulent flows also exhibit drag-reduction\cite{virk1975drag,white2008mechanics} which has been been a subject of intense study.\cite{zakin1998surfactant,li2012turbulent}

Research on flows of WLM solutions has benefited from the vast usage of these complex fluids for many industrial applications\cite{yang2002viscoelastic} which exploit its features including shear-thinning, viscoelasticity, drag-reduction and self-healing. WLM solutions have also been actively employed in the oil-gas industry.\cite{ezrahi2006properties} In this context, a particular type of WLM solution has found to be useful for specific operations.\cite{kefi2004expanding,sullivan2007oilfield} These WLM solutions are made from surfactant molecules that crucially differ from more conventional surfactants in the length of the carbon-chain, consisting of 22 C atoms instead of the more common 12,16 or 18 C atoms. This difference in chemistry has drastic consequences on the rheology of solutions containing these \textit{long-chained} surfactants.

Studies pioneered by Raghavan,\cite{raghavan2001highly,kumar2007wormlike} Feng\cite{chu2010wormlike,chu2011thermo,han2011wormlike} and co-workers have shown that long-chained surfactants in aqueous solutions tend to form elastic gels rather than viscoelastic liquids. This odd rheological behaviour in the absence of physical or chemical cross-links was rationalised to arise from topological entanglements in the presence of worms which are long, temporally persistent, and have a relatively long persistent length.\cite{raghavan2012conundrum} For this paper, our system under investigation is a WLM gel formed by these types long-chained surfactants.\cite{kumar2007wormlike,gupta2021rheology}

While there have been studies on the self-assembly and rheology of WLM gels in the last two decades, few studies have focussed on the flow dynamics of these complex materials. Previous studies from our group have documented the general gel-formation tendencies of long-chained surfactants and their turbulent drag reduction features.\cite{goyal2017,mitishita2022turbulent} More recently, it was shown that WLM gels show evidence of interesting features like long-lived transient shear-banding and shear-induced fracture.\cite{gupta2021rheology}. More generally, the highly elastic nature of WLM gels leads to differences in behaviour when compared to the flow of more conventional WLM solutions.\cite{beaumont2013,mccauley2023evolution} Thus, WLM gels must be treated as a model system whose rheology and behaviour in different flows deserves further investigation.\cite{gupta2023rheology} Having systematically studied the linear and non-linear rheology (using a rheometer) of wormlike micellar gels formed by long-chained zwitterionic surfactants in an earlier paper,\cite{gupta2021rheology} here we focus on the behaviour of these gels when they flow in narrow, round capillaries.  

Flow in capillaries is a canonical problem that appears in some form in a variety of biological, industrial and geophysical settings. For instance, the flow of blood in vessels, complex fluids in transport shunt tubes and flows in network connecting pores are all examples of flows in confined channels which are approximations of flows in a capillary. On a kinematic level, this flow presents a difference from a simple shear flow where the shear rate, $\dot\gamma$ is constant across the geometry. In pipe flows, $\dot\gamma$ varies across the pipe, taking a maximum value on the boundaries. Probing the behaviour of complex fluids in a capillary pipe flow thus presents a natural step up from rheometric flow scenarios and is an important problem to study in the pursuit of understanding the behaviour of complex fluids in real practical settings. However, the flows of WLM fluids in a circular or rectangular capillaries has been relatively less studied when compared to flow in a simple-shear imposed by a rotational rheometer.

Capillary flow has been exploited to calculate rheology of WLM solutions \cite{hernandez1999capillary,salipante2020entrance,murphy2020capillary} as well as understand interesting phenomenon like entry effects,\cite{hashimoto2006effects,salipante2018flow} flow fluctuations\cite{salipante2018flow} and jetting.\cite{haward2014spatiotemporal,salipante2017jetting} Some attention has been directed towards understanding flow profiles and related banding scenario in these geometries. The velocity profiles across the gap for WLM fluids depends on which region of the flow curve is probed by changing a mean $\dot\gamma$. Plug-like profiles gradually develop a shear-band close to the pipe wall in time, when the stress-plateau and following high-shear rate branch are probed respectively. \cite{mair1997shear,britton1999transition,mendez2003particle}Studies conducted in highly confined micro-channel geometries have uncovered dynamics of WLM solutions in such flows and support the existence of shear-banded profiles in such geometries marked by a plug with corresponding thin-layer of high shear-rate.\cite{masselon2008nonlocal,nghe2008high,ober2011spatially,lutz2017situ} The flow of highly elastic wormlike micellar gels in a capillary geometry hasn't been probed to the best of our knowledge and this strongly motivates the current study of the flow dynamics of our wormlike micellar gels in a O(mm) size circular capillary. 

Our main aim is to quantify the types of velocity profiles generated in the flow of these materials, and observe if and how shear banding manifests in the pipe flow of these WLM gels. We focus mostly on transient dynamics which are important in their own right. Not only do they inform the development process to an eventual study state, they also highlight the coupling between rheology and flow dynamics.\cite{kim2016transient} Understanding the feedback between rheological properties and flow dynamics of complex fluids is critical in understanding the various process in which such fluids are employed. As WLM solutions are widely used, there is a value in designing WLM solutions that can be tuned for specific purposes and can respond actively or passively to stimuli.\cite{ketner2007simple,chu2013smart}Not only have WLM gels been synthesized for various applications like tissue engineering\cite{lee2001hydrogels} and nanomaterial fabrication,\cite{bhattacharya2011surfactants,xie2017unique}the genesis of gel formation in these systems bears resemblance to a class of molecular gels\cite{raghavan2012conundrum,raghavan2009distinct} like entangled F-Actin.\cite{mackintosh1997actin} Thus, we expect that our investigation into the capillary flow of WLM gels can aid in understanding flow dynamics of a broader class of complex soft matter and ultimately inform the conception and design of processes that crucially depend on the flow of such materials. 

\section{Materials and Methods}\label{methods}
In this section, we describe the preparation method for the WLM gel used in this paper. We also detail the experimental setup and procedures used to study and characterize the fluid.

\subsection{Sample Preparation}\label{sample}
Our model system is a drag-reducing gravel-packing surfactant product called J590 that is supplied to our laboratory by Schlumberger in liquid form\cite{goyal2017} which we use as is. %Commercial VES products also typically contain a small amount of Sodium Chloride. %
The surfactant solutions contain a mixture of \textit{erucic amidopropyl dimethyl betaine} and propan-2-ol. The former is a zwitterionic surfactant referred to as EAPB\cite{mccoy2016structural} and EDAB\cite{kumar2007wormlike,beaumont2013} in previous studies. We refer to the resultant WLM solution as either VES or WLM gels throughout the paper and expect our results to be applicable to EDAB/EAPB. The concentration of surfactant (by weight or volume) in the WLM gels, ($c$), is only proportional to the \textit{true} surfactant concentration because the exact percentage of surfactant in the provided commercial product is unknown. To prepare the WLM gel, we weigh out an appropriate amount of the provided solution and tap water based on the concentration of the desired WLM gel and mix it vigorously using a mixer (IKA Control 60) at 2000 rpm for 20 minutes, with a high shear rate impeller. The mixed foamy solution is then centrifuged at 3000 rpm for ten minutes by placing it in individual Falcon tubes. Finally, the tubes are placed in a heat bath at temperature of 80$^{\circ}$C for $\sim$ 30 minutes. The final solution i.e. WLM gel is optically transparent. 

For particle based velocimetry purposes, the solutions of prepared complex fluids must be seeded with particles. As seeding particles, we use polystyrene latex beads (Magsphere Inc.) with a monosize diameter, $d_{p} = 4.1\mu$m. These particles are dispersed in a liquid with the solid phase concentration being $\sim 10\%$ by volume and the polymer density of particles, $\rho_{p} $= 1.05 g/ml. We use 50 $\mu$L of this particle-liquid suspension to seed $100$ ml of test fluid. Before seeding, the particle-liquid suspension is vigorously shaken and placed in a sonicator for 30 seconds. After seeding, the test fluid is again placed in a sonicator, degassed and finally centrifuged to prepare a test solution free of particle agglomerates and bubbles.

\subsection{Rheometry}\label{rheometry}
Rheological studies were performed with Malvern Kinexus Ultra+ rheometer and a cup-bob geometry. The inner surface of the cup and the outer surface of the bob are roughened. A solvent trap was used to prevent evaporation of the sample ensuring that evaporation doesn't play a significant role, even for very long tests. Prior to beginning experiments, the fluid was left to stabilize at a desired temperature for $\sim$ 20 minutes. For small amplitude oscillation sweeps (SAOS), we applied a pre-shear at 100 \textit{s}$^{-1}$ for 6 minutes, followed by a resting time of 30 minutes at 0 Pa before every experiment to erase memory of loading and ensure a well defined initial state. After this, the rheometer's software finds an appropriate deformation in the linear viscoelastic region for the material and runs a frequency down-sweep from 3-0.01 Hz. For flow-curves, we conduct a stress-controlled ramp-up from low values of stress (low shear-rate) to a high value of stress (high shear-rate). At each value of stress $\sigma$, the waiting time is 30 minutes and the software averages over the last 10 seconds to report the value of shear-rate $\dot{\gamma}$ for that particular $\sigma$.

\subsection{Capillary Flow Set-up}
For studying the flow in capillary pipes of the fluids described in \ref{sample} we use a capillary tube made of borosilicate glass (Friedrich \& Dimmrock Ltd.) of length, $L$ = 960 mm. The inner diameter of the pipe is $D$ = 1.312 mm.\footnote{To measure the diameter of the pipe accurately we pumped ethanol solution of known refractive index ($\alpha$) through the pipe and imaged the cross-section with the OCT. After identifying the wall edges, the width of the channel is obtained in pixel units ($W$). The final $D$ is then simply $Wh_{pixel}/\alpha$, where $h_{pixel}$ is the height of 1 pixel in $\mu$m.}The capillary tube is enclosed in a triangular prism whose flat side is exploited for imaging purposes and exposed to light when velocity measurements are made, negating the need for constructing an enclosing box around the measurement area to account for curvature of the inner circular capillary. The pipe is mounted on 3-D printed supports that are bolted on a bread-board type rig and then connected to an inline pressure sensor (Elveflow) that is used to effectively measure the pressure difference, $\Delta p$ across the pipe.\footnote{Note that this $\Delta p$ doesn't strictly correspond only to the contributions of flow in the pipe but also in short tubing that connects the syringe to the sensor and the sensor to the pipe (via the manifold). While we have tried to minimize these effects by making sure the tubing length is very small compared to the pipe we cannot, at the level of our experiments, measure the associated $\Delta p$ due to these connections.}To connect the outer triangular prism surface to circular tubings we use a 3-D printed manifold that functions as a converter. The tubing is finally connected to a 10 ml glass syringe (Hamilton Inc.) mounted on a syringe pump (KDscientific). This capillary pipe+board setup is then screwed on-to the base plate of the Optical Coherence Tomography (OCT) device (Thorlabs - TEL1300V2-BU).
\begin{figure}[htb!]
\centering
  \includegraphics[width=0.39\textwidth]{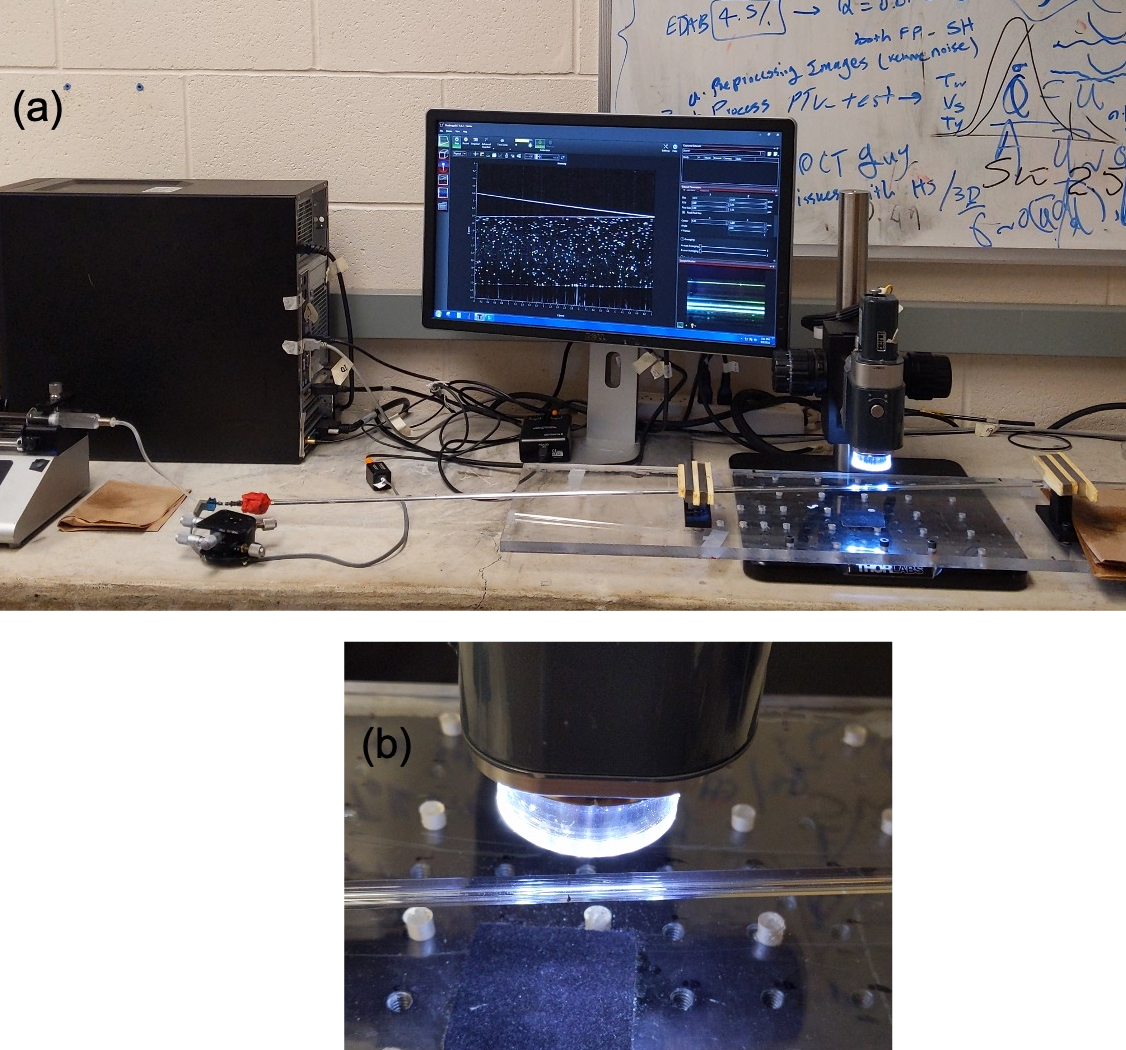}
  \caption{Images of the capillary flow set-up used for experiments. All the elements are clearly seen in (a). (b) shows a zoomed view of the OCT focusing on the test section of the pipe.}
  \label{just}
\end{figure}
\begin{figure}[htb!]
\centering
  \includegraphics[width=0.4\textwidth]{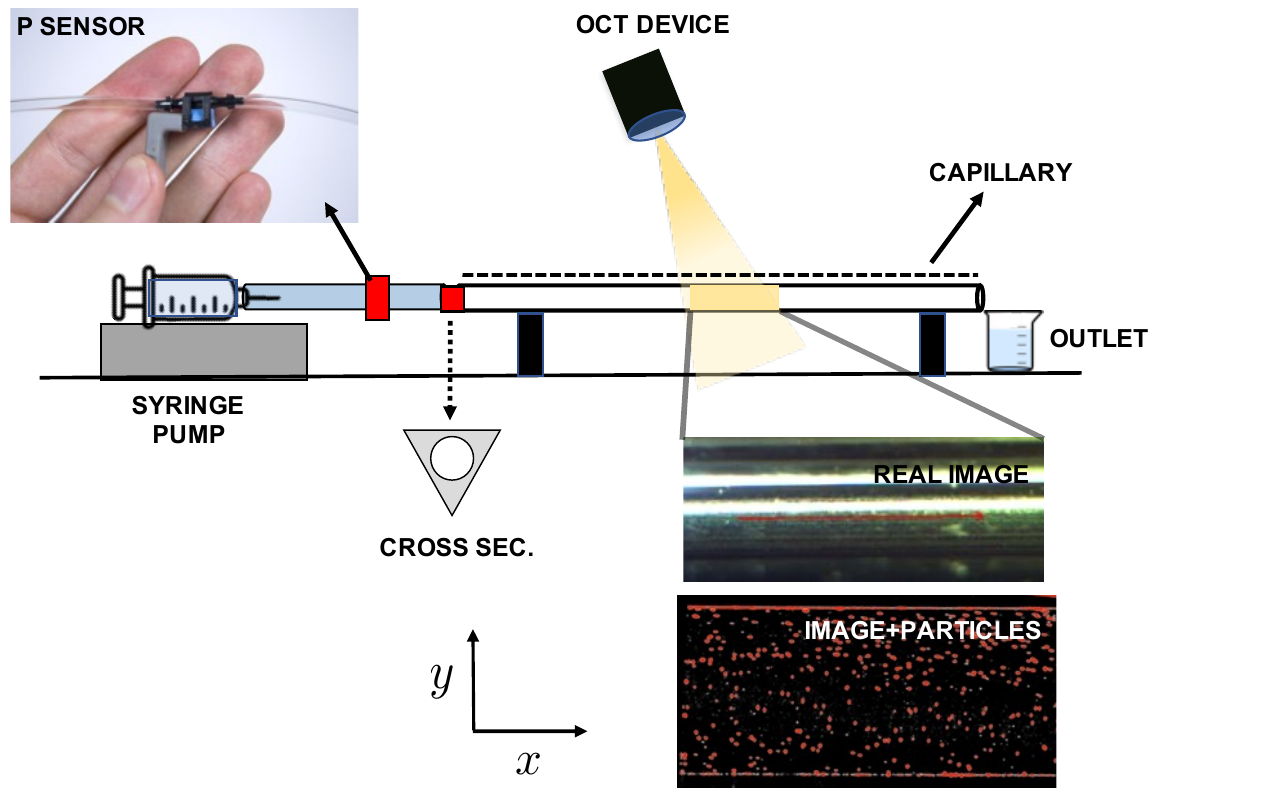}
  \caption{A schematic of the capillary flow set-up used in this paper. Note that the OCT and pipe are kept at an angle $\approx 3^{\circ}$.}
  \label{just}
\end{figure}

The OCT is a real-time imaging modality that works based on interferometry with a broad bandwidth light source. The device is used to obtain cross-sectional images of a specimen from the backscattered light, with micrometer-scale axial and transverse resolution and millisecond temporal resolution\cite{oct} by focusing a collimated beam with the center wave length of 1300 nm on the specimen by using a 5x-objective lens. %The depth of focus of the vertical imaging is approximately 3.5 mm in a medium with the same refractive index as air, while the lateral field of view is adjusted to be 5\;mm. The lateral scanning of the sample is performed by moving {\color{black}the} vertical scanning beam laterally through a two-galvo mirror system. The spacial resolution of vertical imaging of the sample is around 3.5 $\mu$m while its lateral resolution, which is limited to the thickness of the scanning beam, is around 13 $\mu$m.  
Effectively, the OCT allows us to obtain successive images of a cross-sectional area of the flowing sample. The temporal resolution is tuned by adjusting the timestep between successive images, $\Delta t$, which can be set by accessing the various sensitivity modes of functionality (higher sensitivity $\rightarrow$ higher $\Delta t$) as well as changing the lateral field of view or pixel size in the flow direction. Most of the results in this paper were obtained with $\Delta t \in$(10-150) ms. %Re-write OCt
\subsection{Velocimetry and Particle Tracking}
To perform Particle Imaging Velocimetry (PIV) and Particle Tracking Velocimetry (PTV) we use the DaVIS software (DaVIS 8.4.0, LAVision). For PIV we use a multi-pass iteration with the size of the larger interrogation window being 64/48 and the size of the smaller window being 32/24/16 in pixel units. The overlap percentages for the larger and smaller window is fixed at 50 and 75\% respectively. The spatial resolution of PIV is of the order of the size of the interrogation window which for a pixel size of 3.53 $\mu$m gives us a resolution between 56-112 $\mu$m. When a higher spatial resolution for velocity profiles is desired we switch to using the PIV+PTV mode of the software which functions as an integrated method where PIV is performed as a first step and the results are further refined by particle tracking in the next step.\cite{scarano2000advances,theunissen2004novel,stanislas2005main} Ideally, PTV provides a resolution of size $d_{p}$ but in-practice, owing to imperfect imaging (especially near the walls) etc, we get spatial resolution $\sim$ O($d_{p}$). Note, that even when the resolution of PTV near the walls is less than ideal, PTV is still better suited than PIV to resolve sharp velocity gradients.

For constructing profiles for streamwise velocity ($u_x(y)$) at a single location ($x$ coordinate) along the pipe we use a field of view of approximately 5mm. Vectors lying in this field of view are binned based on their $y$ coordinate. For each bin the median is picked as a representative velocity at that $y$ coordinate for that time $t$. At $t$ then, we get a single velocity profile $u_x(y) = u$ across the channel. To construct velocity time series we repeat this spatial binning for multiple frames. For time-averaged PIV profiles shown later in the paper, we simply do a mean of all profiles over a time period. For time-averaged PTV profiles, we repeat the binning procedure but now the binning process is time agnostic. Note, that we define the \textit{plug} region as the part of the velocity profile where shear-rate $\dot\gamma = 0$. For all velocity profiles presented in this paper, we plot $u$ in mm/s and $y$ coordinate is scaled with $D/2$, thus $y$=1/-1 correspond to the top/bottom wall respectively. For calculating particle trajectories, we load images in ImageJ software and use its native particle tracking algorithm (Mosaic\cite{mosaic}) to track individual particles. Appropriate particle trajectories are chosen by considering the length of the trajectory (to garner enough data) as well as location of particle (for eg. near the wall, in the shear-layer).

\subsubsection{Experimental Procedure}

Before running every flow test, we flush the capillary pipe with ethanol and then water at a flow rate of 5 ml/min. This adequately prepares the pipe for the test and ensures that particles stuck in the pressure sensor or the walls of the pipe will be cleared away. At the start of the test, we load the syringe with the sample fluid and mount it on the syringe pump. When working with VES samples we let the sample rest in the syringe for $\sim$ 10 minutes before starting the flow. At $t=0$, we also start recording $\Delta p$ vs $t$. %The end of the test is typically when the syringe is emptied. %  
All measurements with VES solutions are made at a distance of 700mm from the pipe inlet, whereas profiles reported in Appendix A\nameref{benchmark} are made at 500 mm from the inlet. For VES solutions, we work with three solutions of different concentrations, $c$ = 1\%,2\% and 3\% at four values of imposed flowrate, $Q= $ 0.05,0.1,0.25,0.5 ml/min. Besides \textit{start-up} tests in which we simply pump the fluid at an a fixed $Q$, we also run \textit{step-down} tests (reported in Appendix B\nameref{stepdown})  where we impose $Q_{upper}$ = 3 ml/min for 2 minutes followed by a $Q_{lower}$ = 0.05 or 0.25 ml/min and \textit{flow-cessation} tests in which flowrate is abruptly set to 0 after imposing a fixed flowrate for a period of time.

Before studying the flow dynamics of VES in our capillary system we ran a few benchmark tests to check for proper working of our experimental set-up and robustness of methods and techniques employed. For this purpose we calculated the viscosity of standard glycerol solution to benchmark the pressure sensor and quantified velocity profiles for glycerol, PEO solution (viscoelastic fluid) and Carbopol solutions (viscoplastic gel with yield stress). The results of these tests are summarised in Appendix A.\nameref{benchmark}

\section{Rheology of wormlike micellar gels}\label{rheology}

\begin{figure}[h!]
\centering
\hspace{-0.7cm}
  \includegraphics[width=0.43\textwidth]{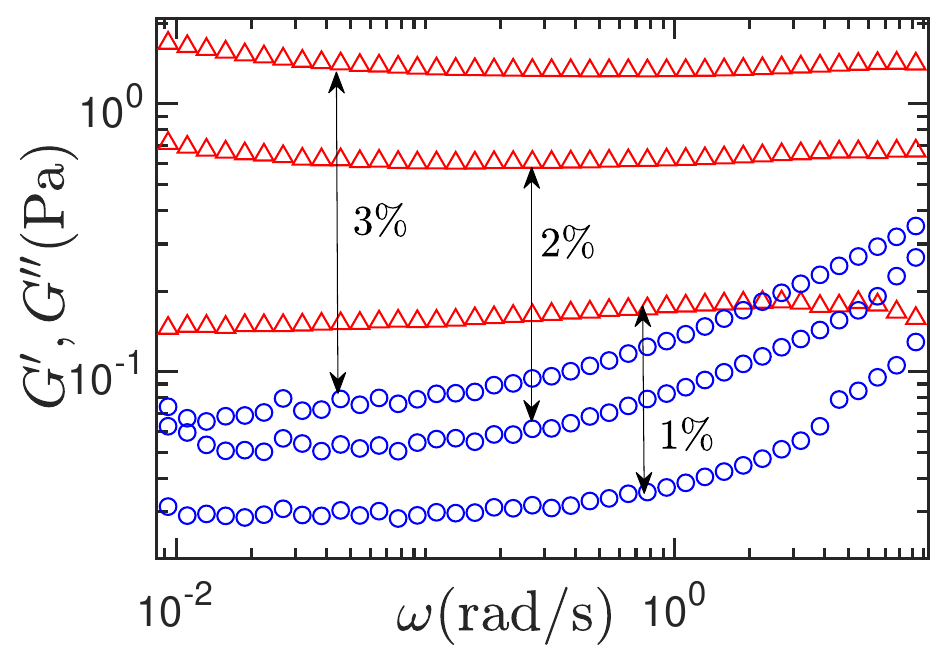}
  \caption{: $G'$ (Red Triangles) and $G''$ (Blue Circles) vs $\omega$ for 1,2,3\% VES solutions from a SAOS test.}
  \label{saos}
\end{figure}

Here, we report the results of SAOS and shear-ramp tests for the wormlike micellar gels formed by 1-3\% VES solutions which inform the gel nature of the WLM solution and inform our capillary flow experiments. In Fig.~\ref{saos} we see that for more than two decades in frequency $\omega$, the storage modulus $G'$ is independent of $\omega$, whereas the loss modulus $G''$ is independent of $\omega$ in the low $\omega$ regime while having a power-law like dependence for higher $\omega$, reminiscent of the anomalous increase in $G''$ \cite{liu1996anomalous} with $\omega$ reported for Carbopol. \cite{divoux2011stress,conley2019relationship,migliozzi2020investigation} The lack of a crossover in $G'$ and $G''$ in the range of $\omega$ probed indicates very long structural relaxation times \cite{gupta2021rheology,raghavan2001highly,kumar2007wormlike} and the fact that values of $G'$ are larger than $G''$ by almost an order of magnitude (atleast for 2,3\% solutions) motivates us to refer to these VES solutions as wormlike micellar gels. Note, that these wormlike micellar gels exhibit a sort of gel-sol transition with temperature, where at higher temperatures, their linear rheology is that of a Maxwell-type viscoelastic fluid with a single relaxation time.\cite{gupta2021rheology}Experiments performed with VES in this paper were run at $T = 24-25^{\circ}$, well below the gel-sol transition.

\begin{figure}[htb!]
\centering
\hspace{-0.7cm}
  \includegraphics[width=0.42\textwidth]{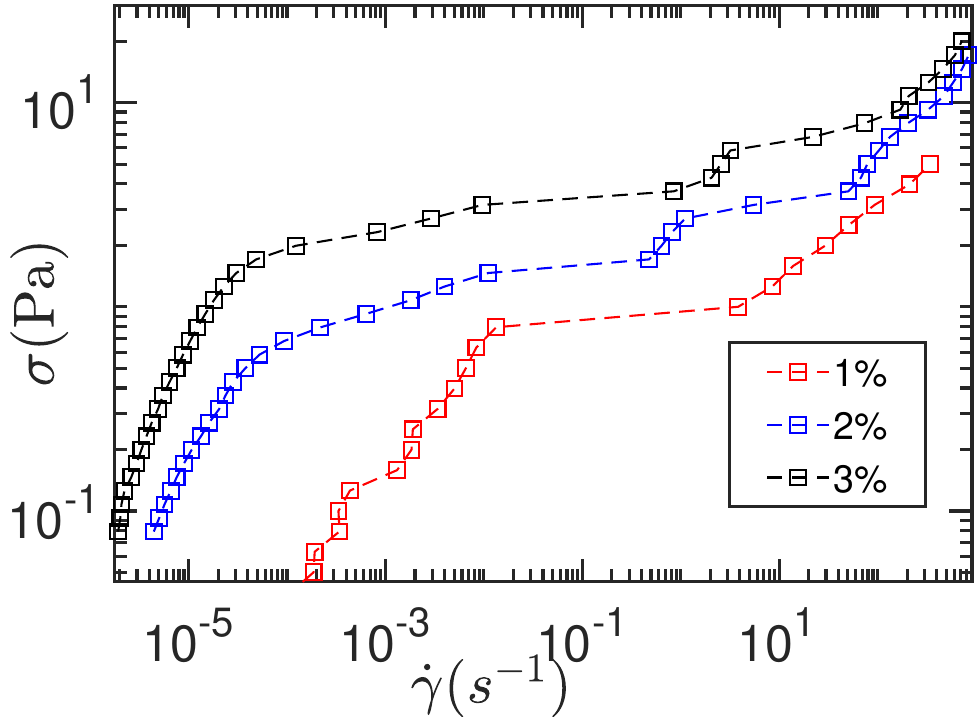}
  \caption{Transient flow-curves ($\sigma$ vs $\dot\gamma$ for 1,2,3\% VES solutions).}
  \label{flowcurve}
\end{figure}

As mentioned in \ref{rheometry}, the flow-curves in Fig.~\ref{flowcurve} are produced by a ramp-up of imposed stress $\sigma$ with each step held for a fixed time. The flow-curves exhibit two shear-thinning regions separated by a plateau(s) in shear stress. Note that a slope for the stress plateau is expected for Couette systems due to stress inhomogenity.\cite{radulescu2000matched,salmon2003shear} In the case of 2\% and 3\% VES solutions there are 2 ramps separated by a small thinning region. We don't have an explanation for this, but this feature is repeatable and is likely dependent on the ramp kinetics employed. Owing to the long timescales involved in the flow of these materials, \cite{gupta2021rheology} these flow-curves are sensitive to the waiting time per step in the ramp and thus the value of $\sigma$ is only an approximation of the steady state $\sigma$ at that $\dot{\gamma}$. In that light, the shear-stress plateau observed in the flow-curves in Fig.\ref{flowcurve} must be interpreted as a sign of possible \textit{transient} shear-banding.
In the absence of better resolved measurements, we cannot comment on if these bands persist in steady-state, but as mentioned in our previous work,\cite{gupta2021rheology}its likely that entanglement kinetics \cite{hu2005kinetics} leads to shear-banding in our wormlike micellar gel system at some level, possibly in a mechanism akin to that proposed for soft glassy matter. \cite{martin2012transient,mccauley2023evolution} The stress plateau is followed by the high-shear rate branch that, for conventional surfactant solutions, is usually only accessible via capillary rheometry. But, in the case of our wormlike micellar gels, begins at values of $\dot\gamma$ already accessible by rotational rheometry (Couette), presumably owing to the very long relaxation times in our system. The long relaxation time is also responsible in pushing the Newtonian branch of the flow-curve to very low values of $\dot{\gamma}$. We end our discussion of the rheology of VES solutions here and move on the dynamics of these solutions in a capillary pipe flow. The reader is directed to \citeauthor{gupta2021rheology}\cite{gupta2021rheology} for a more detailed discussion of the linear and nonlinear rheology of these specific wormlike micellar gels. 
\section{Results}
In this section we discuss the flow dynamics of wormlike micellar gels in capillary pipe flow by presenting the evolution of velocity profiles in time for different surfactant concentrations, $c$ and flowrates, $Q$. We also discuss flow behaviour flow-cessation tests and rationalise these results in the context of the rheology of wormlike micellar gels. 
\subsection{Plug flows in 3\% VES solutions}\label{3percent}
\begin{figure}[htb!]
\centering
  \includegraphics[width=0.42\textwidth]{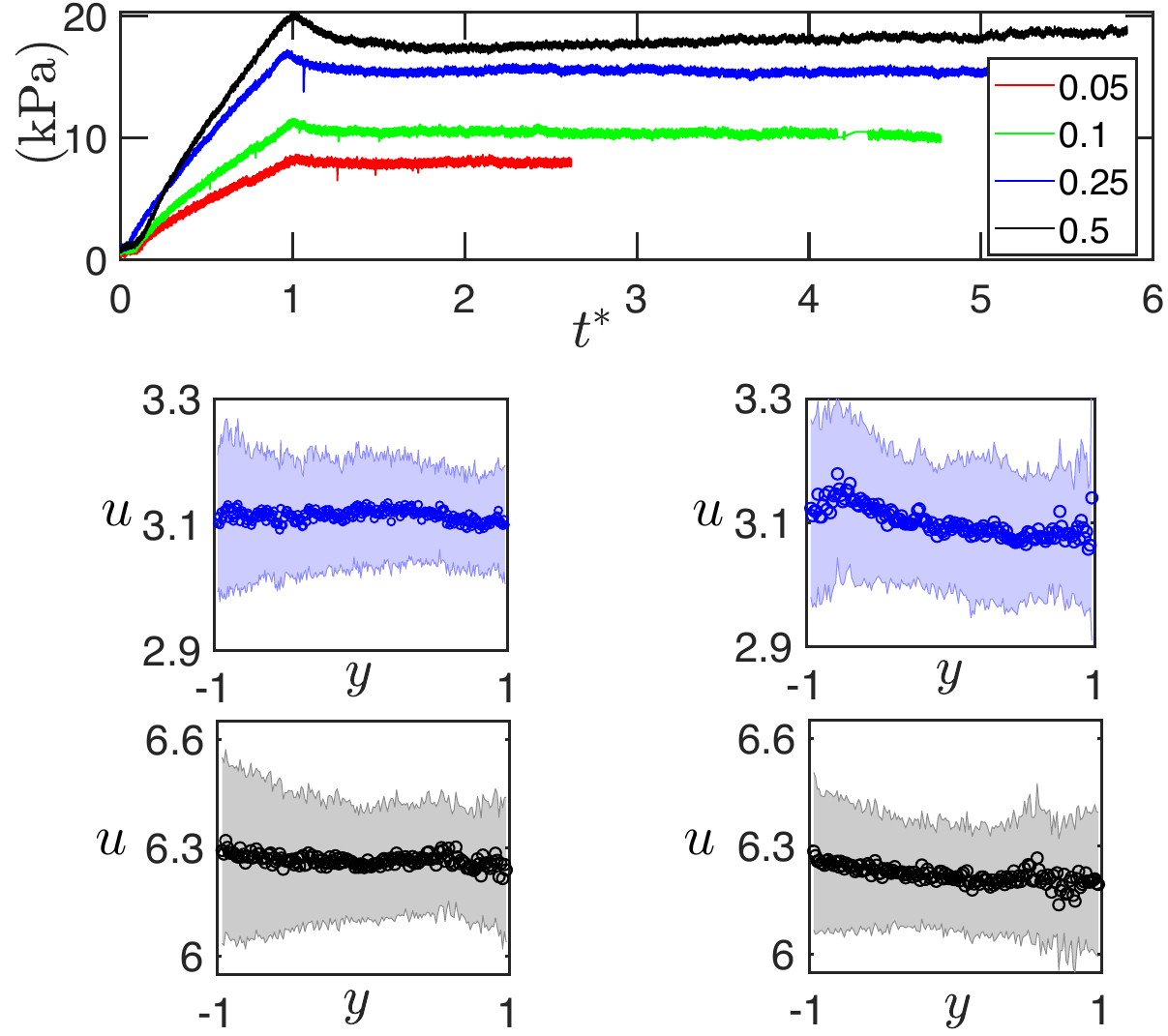} 
  \caption{Dynamics for start-up flow of 3\% VES. Top : $\Delta p$ vs $t^* = t/t_f$  for $Q$ = 0.05,0.1,0.25,0.5 ml/min, Middle :  Early (left) and late-time (right) $u$ vs $y$ profiles calculated by PTV for $Q$ = 0.25, Bottom : Early (left) and late-time (right) $u$ vs $y$ profiles calculated by PTV for $Q$ = 0.5.}
  \label{3plugvel}
\end{figure}

We start this section by reporting the results for the flow of the most concentrated VES solution employed in this study - 3\%. In the $\Delta p$ time series for different values of $Q$ in Fig.~\ref{3plugvel}\footnote{In Fig.~\ref{3plugvel} and Fig.~\ref{2plugp} we have subtracted out the initial time it takes to fill the connected tube to get the peaks in $\Delta p$ corresponding to the time period, $t_f$ coincide at 1 for different tests.}, we can identify regions that correspond to the pipe being gradually filled by the wormlike micellar gel by displacing the water left over in the flushing procedure before the start of the test. In this stage, $\Delta p$ increases for a time, $t\approx \frac{L}{\bar U} = t_f$ where $\bar U$ is a mean velocity that is equal to $\frac{4Q}{\pi d^2}$. This increase in $\Delta p$ corresponds to the dissipation induced by this fluid displacement. In Fig.~\ref{3plugvel}, we see that after the initial increase, the value of $\Delta p$ first slightly drops. The black curve in Fig.~\ref{3plugvel}(Top) also seems to rise slightly in time. These are likely due to some time-dependent change in structure. After the slight drop, the $\Delta p$ curves saturates to a steady value that persists till the end of the test. This steady value of $\Delta p$ increases with increasing flowrate. We will discuss the interpretations of these $\Delta p$ values later in the section and instead discuss the velocity profiles for the flow of 3\% VES solution in our capillary pipe.
In Fig.~\ref{3plugvel}, we plot the time-averaged velocity measured at early and late times in the test for $Q$ = 0.25,0.5 ml/min. The profiles show a wall to wall plug flow that is stable in time. This behaviour is also present for the other values of $Q$ (not shown). We have also carried out limited measurements for higher concentrations of VES (upto 8\%) and have found the same, i.e the flow remains a stable plug at these flowrates. We call this a \textit{full} plug because our velocity measurements detects no presence of a shear layer. This implies that the wormlike micellar gel, at this concentration, essentially flows like a solid plug with strong slip at the wall. Note that even if calculated velocity profiles like those in Fig.~\ref{3plugvel} have their closest data point $\sim$ 20$\mu$m from the wall, we have visually checked particle trajectories nearer in videos and can't detect any visible deviation from the plug velocity (except in very short-lived transients). 

Typically, wormlike micellar solutions show a Newtonian parabolic velocity profile for a mean $\dot{\gamma}$ in the lower branch of the flow-curve.\cite{britton1999transition,mendez2003particle,mair1997shear} As discussed in Section.~\ref{rheology}, this branch occurs for very low values of $\dot{\gamma}$ which are inaccessible in this capillary flow study. We can define a mean shear-rate, $\bar{\dot{\gamma}}  \textcolor{red}{=} \frac{8\bar U}{d}$, which for the values of $Q$ employed in this study is orders of magnitude larger than the $\dot{\gamma}$ in the eventual Newtonian region of the flow-curve. Therefore we expect to not see parabolic type profiles for the flow of VES solutions in our capillary tube. A possible way to rationalize this plug flow with strong wall slip is by assuming that the plug is lubricated by a layer of surfactant free solvent (water) with viscosity, $\eta_w$. This layer formation is mediated by a flow-induced desorption effect that is expected in pipes without surface treatment,\cite{hemminger2017microscopic} like untreated glass used in this study. This mode of wall slip has often been invoked for strong slip observed in polymeric fluids, \cite{de1981polymer,cuenca2013submicron} in lamellar phase surfactant solutions \cite{salmon2003shear} and branched  \cite{caiazza2019flow} and in semi-dilute \cite{decruppe2006local} wormlike micellar solutions. If the lubricating layer is made up mostly of the solvent then the wall shear stress, $\sigma_w$ can be approximated by the formula $\sigma_w \approx \eta_w\frac{u_{slip}}{\delta}$ where $u_{slip}$ and $\delta$ are the slip velocity and lubricating layer thickness and $\eta_s$ is the shear viscosity of water. For $Q$=0.5 ml/min, we have a mean $\Delta p$ = 18.3 kPa. Using $\sigma_w = \frac{D\Delta p}{4L}$, where $\sigma_w$ is the wall shear stress we get $\sigma_w$ = 6.27 Pa. $u_{slip}$ for cases with full plug is similar to $\bar U$ at that $Q$. Combining this information, we get $\delta \approx$ 1 $\mu$m. This is lower than the maximum resolution of our velocimetry, i.e $\approx$ 4 $\mu$m. If we repeat the same calculation for $Q$ = 0.05 ml/min we get $\delta \approx$ 0.2 $\mu$m, which is also less than our velocimetry resolution. \begin{comment}
Note, that if indeed this is the mechanism for the observed wall-slip, then these values of $\Delta p$ should closely reflect the true values of pressure drop for the flow as the development length associated with such depletion layer is not expected to be very large. This is one of the reasons why Carbopol solutions which have a similar lubricating layer, don't have large development lengths associated with their pressure or velocity. 
\end{comment}

One of the other ways to think about the unresolved layer of fluid that lubricates the plug could be in the context of shear banding. As mentioned in Section.~\ref{rheology} and in our previous study \cite{gupta2021rheology}, our wormlike micellar gels show signatures of shear banding. A rough estimate of the thickness, $l_{sb}$ of the shear band can be obtained by using the formula, $\dot{\gamma_u} \approx \frac{u_{slip}}{l_{sb}}$ where $\dot\gamma_u$ is the shear rate at which the shear-thinning upper branch of the flowcurve commences.\cite{ober2011spatially}. It's not straightforward how to pick $\dot\gamma_u$ from a transient flowcurve. We have noticed that in a \textit{down sweep} flow-curve the plateau in shear-stress is shifted towards very low values of $\dot\gamma$.\cite{gupta2021rheology}. For purpose of estimation, suppose that $\dot\gamma_u \sim O(100) s^{-1}$ and pick the lowest value of $u_{slip}$ that occurs at the lowest value of $Q$ i.e. 0.05 ml/min. This gives us an estimation for the shear band thickness for 3\% solution, $l_{sb} \sim 6 \mu$m which is within our experimental resolution. This means that our velocimetry analysis should be able to pick up the presence of a shear layer (especially at higher $Q$ for which $u_{slip}$ and hence $l_{sb}$ increases). Discussion of the velocity and pressure profiles for 2\% VES solutions will help shed some light on the nature of the plug and associated wall slip. 
\subsection{Plugs and shear-layers in flows of 2\% VES solutions}\label{2percent}
\begin{figure}[htb!]
\centering
  \includegraphics[width=0.42\textwidth]{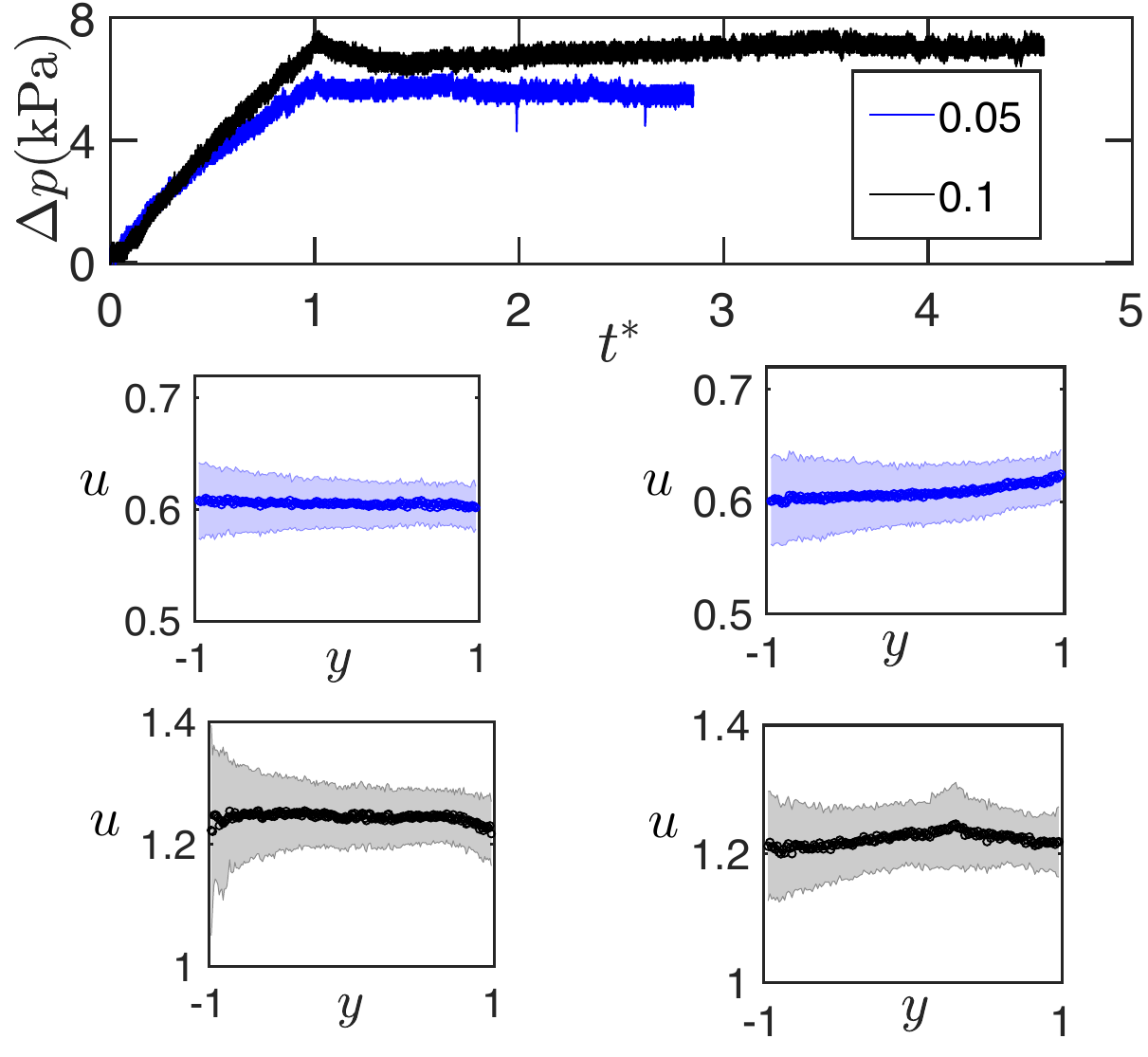}
  \caption{Dynamics for start-up flow of 2\% VES. Top : $\Delta p$ vs $t^* = t/t_f$ for $Q$ = 0.05,0.1 ml/min, Middle :  Early (left) and late-time (right) $u$ vs $y$ profiles calculated by PTV for $Q$ = 0.05 ml/min, Bottom : Early (left) and late-time (right) $u$ vs $y$ profiles calculated by PTV for $Q$ = 0.1 ml/min.}
  \label{2plugp}
\end{figure}

\begin{figure}[htb!]
\centering
  \includegraphics[width=0.42\textwidth]{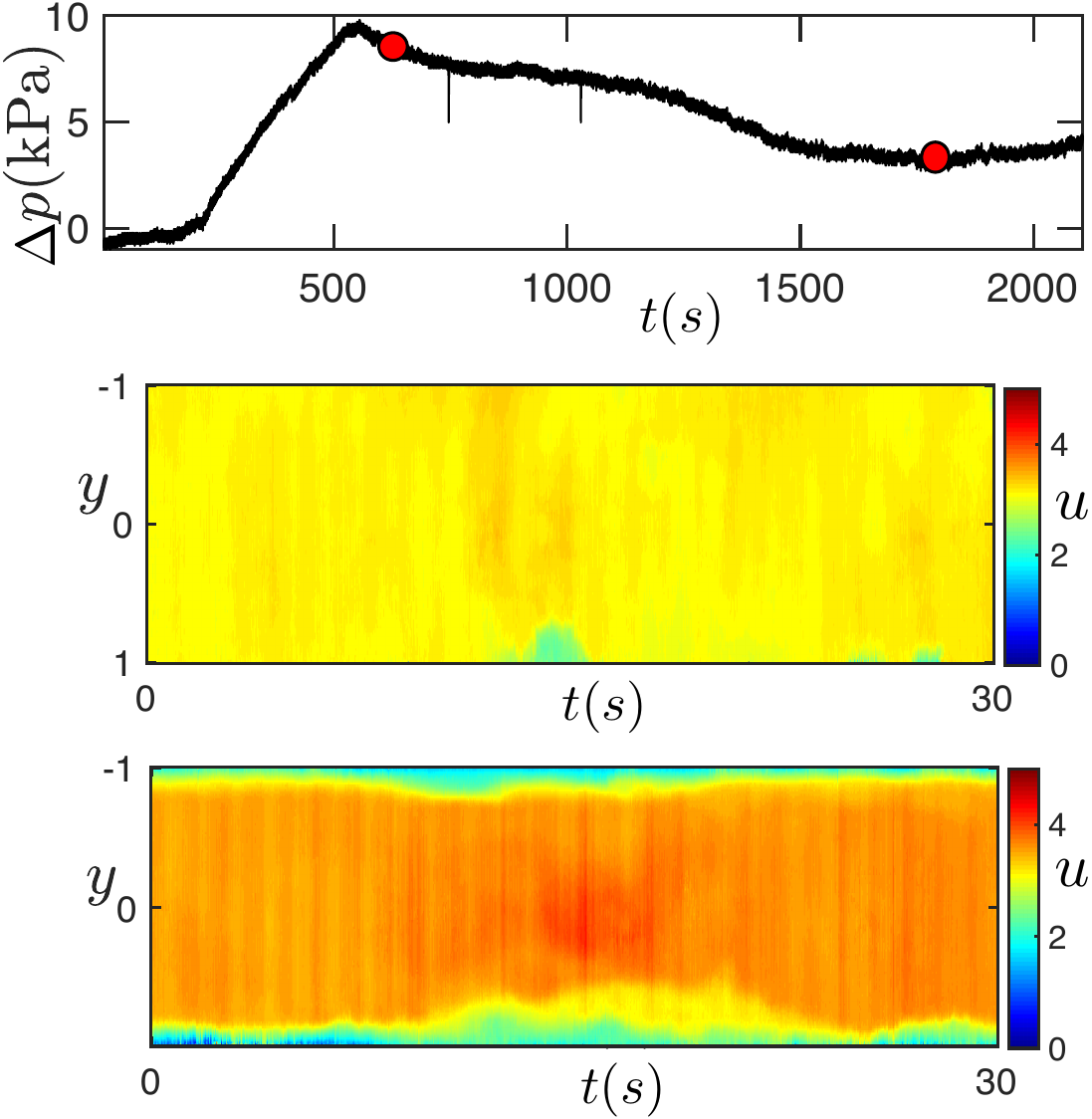}
  \caption{Dynamics for 2\% VES at $Q$ = 0.25 ml/min. Top : $\Delta p$ vs time, Middle : early time velocity time series, Bottom : late time velocity time series. Red circles in the top panel indicate the time from which time series are sourced.}
   \label{2_25}
\end{figure}

In Fig.~\ref{2plugp}, we see that the $\Delta p$ time series for $Q$ = 0.05 and 0.01 ml/min is similar to those plotted in Fig.\ref{3plugvel} i.e there is the initial increase followed by a near saturation. 
The velocity profiles show a plug-flow akin to those observed for more concentrated solutions. Repeating the calculation done for 3\% VES solutions, an estimate for $\delta$ is higher than that reported for 3\% solutions (as $\Delta p$ and thus $\sigma_w$ reduces as $c$ reduces for the same $Q$). An estimate for $l_{sb}$ should be similar if we keep the estimate for $\dot\gamma_u$ the same as before (but likely that $l_{sb}$ is larger as $\dot\gamma_{u}$ for 2\% VES is less than that for 3\% solutions). While we don't detect a shear layer even for 2\% solutions at these flowrates in our experiments, we see in Fig.~\ref{2_25} that for $Q$ = 0.25 ml/min, a shear-layer finally appears. 
In Fig.~\ref{2_25}, we first plot the $\Delta p$ time series for $Q$ = 0.25 ml/min and see that its behaviour is not the same as the $\Delta p$ time series for flows with full-plug. Here after the initial increase in $\Delta p$, there is a gradual drop followed by a region that is almost steady towards the end of the test.  In the middle and lower panels of Fig.\ref{2_25}, we show the time series of velocity evolution in a time period of $\Delta t\sim$ 30s at both an early and late time in the test. At early times, we see that the flow is a full plug similar to the lower flowrates for 2\% solution and the complete range of $Q$ for higher $c$ solutions. In pipe flow, the magnitude of shear stress varies linearly between the centre and wall regions where the fluid experiences maximum shear stress. These near-wall regions also flow at lower velocities and therefore have a longer residence time in the pipe. These two factors allow the accumulation of strain in near-wall regions, leading to progressive alteration of the micellar network near the wall. The last panel in Fig.\ref{2_25} shows that the flow at late-times has a distinct plug region where $\dot\gamma$ = 0, and near-wall region where shear is concentrated. The interface between the two fluctuates in time. These features, namely a large plug region at early times and plug+fluctuating shear-layer at late-times, are also repeated for $Q$ = 0.5 ml/min. The time-dependent evolution of the shear-layer highlights the strongly \textit{transient} nature of the flow of our wormlike micellar gels.
\begin{figure}[htb!]
\centering
  \includegraphics[width=0.42\textwidth]{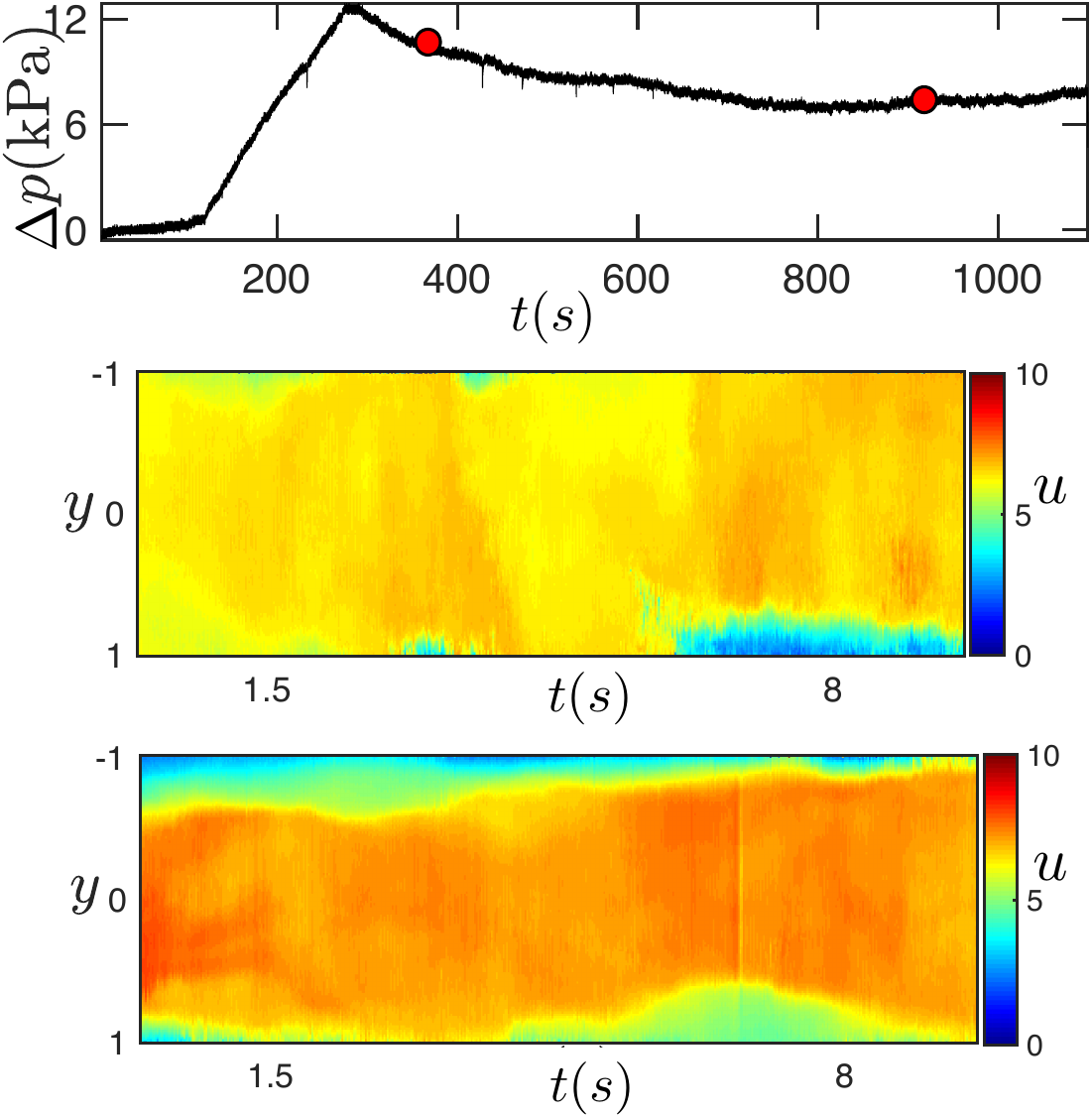}
 \caption{Dynamics for 2\% VES at $Q$ = 0.5 ml/min. Top : $\Delta p$ vs time, Middle : early time velocity time series, Bottom : late time velocity time series. Red circles in the top panel indicate the time from which time series are sourced.}
\label{2_50}
\end{figure}

%We also note that the velocity profiles calculated are averaged over a spatial area of a single plane. As such it is not possible to comment on the azimuthal angular dependence of velocity profiles. But, the flow-rate calculated by integrating the presented velocity profiles typically fluctuates around the mean imposed flowrate, $Q$, thereby making our assumption of axisymmetric flow not entirely unreasonable.\end{comment}

\subsection{Plug Flow as a Developing Flow}
An odd feature of the flow dynamics at $Q$ = 0.25 ml/min for 2\% VES is that if we try and extract a value of $\Delta p$ for the final flow profile (with shear-layer) by averaging the $\Delta p$ time series in the last 2 minutes and compare it with the value of 'average' $\Delta p$ obtained by repeating the operation for $Q$ = 0.1 ml/min, we see that the latter is \textit{higher} than the former. This is counterintuitive as we expect that $\Delta p$ to increase with an increase in imposed flow-rate. For Newtonian fluids, $\Delta p$ linearly grows with $Q$, whereas with shear-thinning fluids, the relationship is sublinear but still increasing function of $Q$. 
%The measure of work done in pumping at a desired $Q$ is obtained by simply calculating the area under the $\Delta P$ time series curve for the same length of time. We have checked that the work-done to pump at $Q$ = 0.25 $\ge$ $Q$ = 0.1 ml/min. However, we observe that this inequality is largely satisfied owing to the $\Delta P$ contributions of the initial increase due to displacement of existent water in the pipe (i.e for $t<t_f$). 
In light of this how do we reconcile the relatively larger values of $Q$ = 0.1 ml/min in comparision with $Q$ = 0.25 ml/min? We see this question posed even in the $\Delta p$ time series for two step-down experiments with 2\% VES solutions (See black curves in Fig.~\ref{step_005} and Fig.~\ref{step_025} in Appendix B\nameref{stepdown}) where the $\Delta p$ curves for $Q$ = 0.05 and 0.25 ml/min almost lie on top of each other after step-down from the larger $Q$ = 3 ml/min.  

We believe that these large values of $\Delta p$ for lower flow rates stems from pressure contributions of the \textit{developing} flow associated at these flow-rates. Wormlike micellar gels have a very large viscoelastic relaxation times, $\lambda \sim O(10^5 s).$ If we define a Deborah number, $De = \bar{U}\lambda/L_{obs}$, for $L_{obs}\sim O(1 m)$, we get $De \sim O(100)$ for $Q$ = 0.05-0.1 ml/min, provided we take the `equilibrium' values of $\lambda$\cite{gupta2021rheology}. This suggests that the elastic stresses haven't fully relaxed in the observational window our pipe length provides, indicating that the flow is transient and developing in this regime.\cite{ober2011spatially} Note that $\Delta p$ comparisons made earlier were done using only the late-time pressures for $Q$ = 0.25 ml/min. Flow at this rate would have a development length and an associated pressure-drop contribution too. But as seen in the time series in Fig.~\ref{2_25}, the pressure drops continuously after the initial rise, not unlike the drop in stress after an overshoot in a shear startup. Thus, in averaging the late-time $\Delta p$ we likely bypass much of the $\Delta p$ contributions coming from the development of flow. Development lengths and associated contributions to $\Delta p$ were systematically reported in Salipante et.al \cite{salipante2020entrance}. In order to characterize the $\Delta p$ values and flow profiles for a fully developed flow we would likely need a much longer pipe (probably $O(De)$ times longer) and carry out long-time experiments. Additionally, multiple shorter pipe experiments would be needed to subtract out contributions of the developing region.\cite{salipante2020entrance}These are out of scope of the present study. 
However, looking at the flow of our VES solutions as a transient developing flow of a viscoelastic fluid helps us rationalize the issues with $\Delta p$ discussed earlier in Section~\ref{3percent}. 

\subsection{Flow Dynamics of 1\% VES and Possible Shear Banding}
 Shear banding in a pipe isn't as widely studied as shear banding in a Couette cell \cite{salmon2003velocity,lopez2004shear,hu2005kinetics,becu2007evidence,decruppe2006local}, where shear-banding manifests as a velocity profiles with two (or more \cite{zhou2012multiple,casanellas2015spatiotemporal}) shear rates characterized by a sharp kink(s) in the velocity profile. The analogue shear-banding velocity profile in confined channel and pipe flows is found to consist of a plug flanked by shear-layers and the transition from zero to finite $\dot\gamma$ is sharp. \cite{mendez2003particle,masselon2008nonlocal,nghe2008high,masselon2010influence,ober2011spatially,lutz2017situ} 
\begin{figure}[htb!]
\centering
  \includegraphics[width=0.42\textwidth]{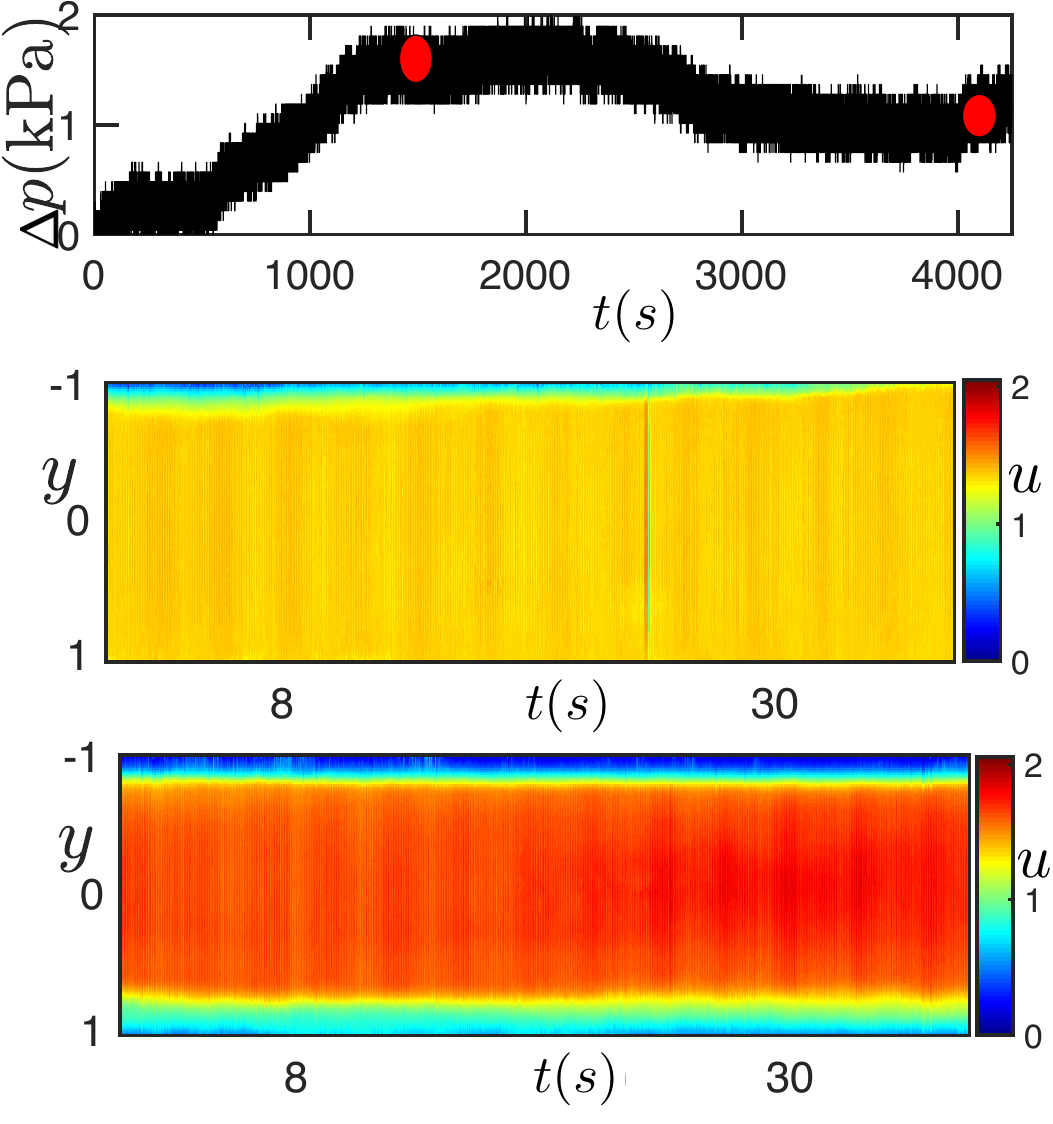}
\caption{Dynamics for 1\% VES at $Q$ = 0.10 ml/min. Top : $\Delta p$ vs time, Middle : early time velocity time series, Bottom : late time velocity time series. Red circles in the top panel indicate the time from which time series are sourced}
\label{1_10}
\end{figure}
\begin{figure}[htb!]
\centering
  \includegraphics[width=0.42\textwidth]{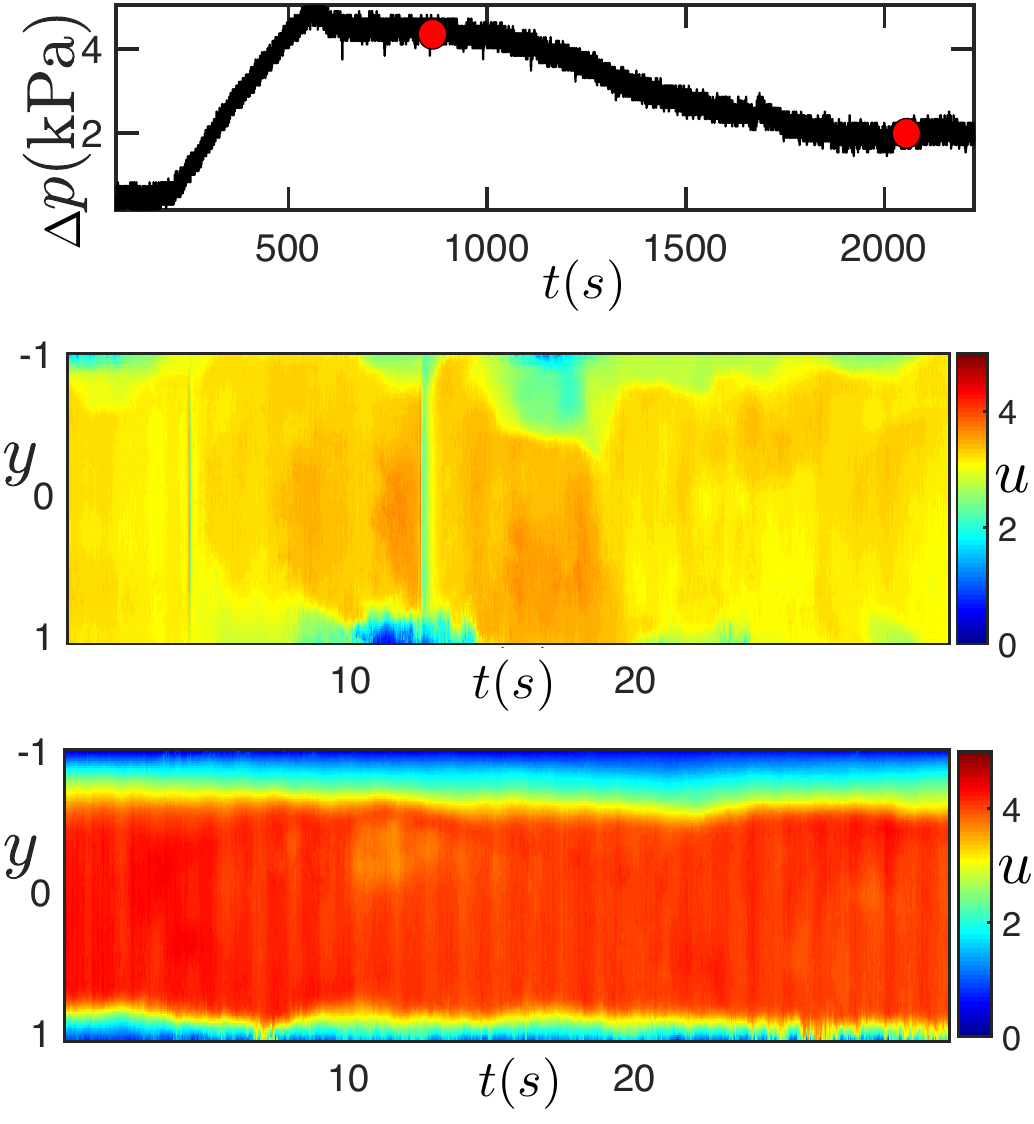}
\caption{Dynamics for 1\% VES at $Q$ = 0.25 ml/min. Top : $\Delta p$ vs time, Middle : early time velocity time series, Bottom : late time velocity time series. Red circles in the top panel indicate the time from which time series are sourced.}
\label{1_25}
\end{figure}

\begin{figure}[htb!]
\centering
  \includegraphics[height=5cm]{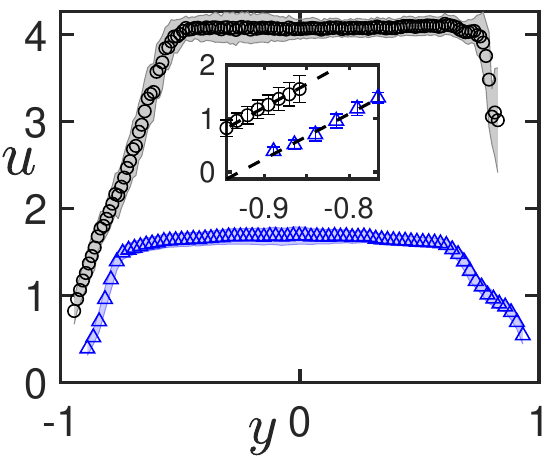}
  \caption{Averaged velocity profiles for 1\% VES at for $Q$ = 0.1ml/min (blue triangles) and $Q$ = 0.25 ml/min (black circles). Inset : Last few points for both main figure curves fit to a straight line (color+symbols same as main figure).}
  \label{1per_ptv}
\end{figure}

\begin{figure}[htb!]
\centering
  \includegraphics[width=0.45\textwidth]{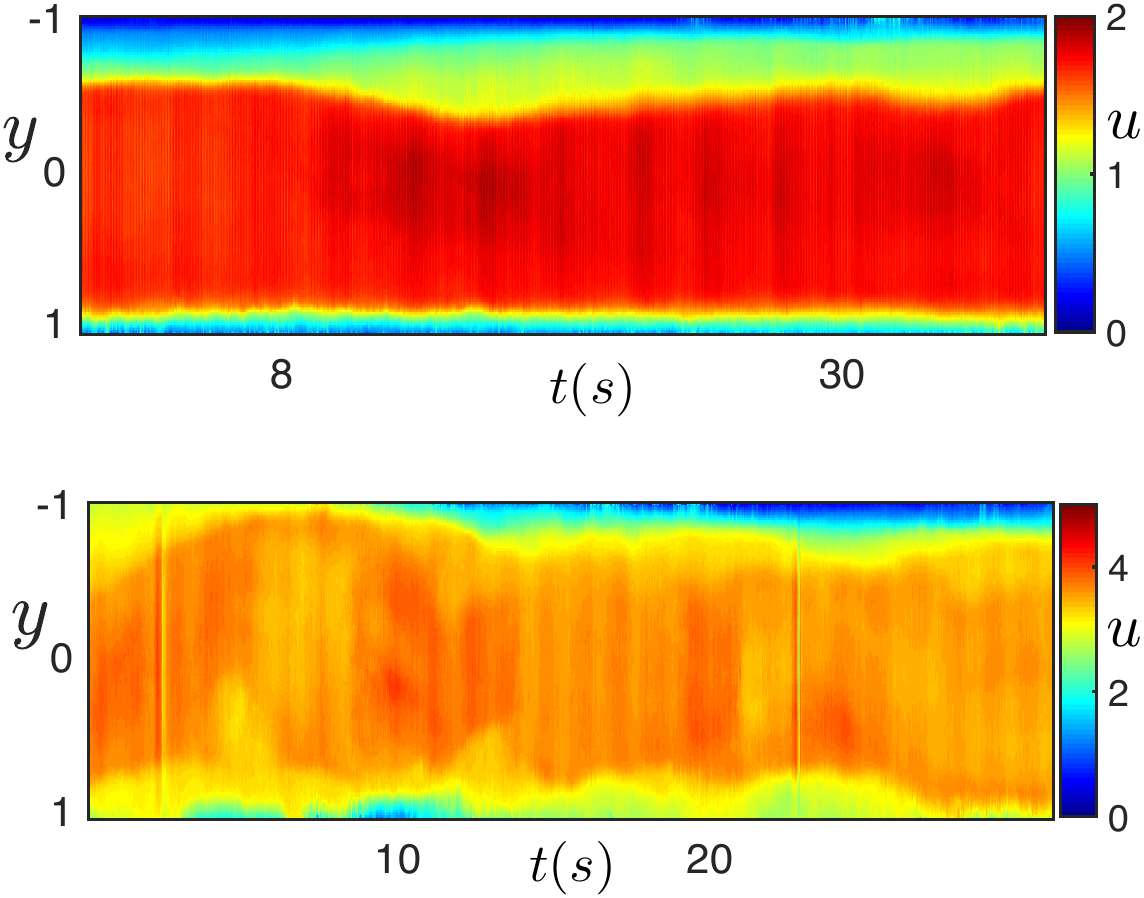}
  \caption{Velocity time series for intermediate times for 1\% at  $Q$ = 0.1 (top) and 0.25 ml/min (bottom).}
\label{1_bet}
\end{figure}

Owing to the highly irregular development of shear layers in the flow of 2\% VES solutions, it is unreasonable to extract a velocity profile in a time-averaged sense. However, we see that for 1\% VES (at $Q$ = 0.1,0.25 ml/min as seen in Figs.~\ref{1_10},\ref{1_25}), the shear-layer is more well-behaved, enabling us to use PTV to compute time-averaged velocity profiles for late times. The early and late-time velocity plots in Figs.~\ref{1_10},\ref{1_25} adhere to the general dynamics we have outlined for WLM gels in this paper, namely that moving forward in time, the flow develops a shear layer (for 1\% VES, the early-time velocity does have a shear layer that \textit{grows} with time). 
We see that the PTV velocity profiles in Fig.\ref{1per_ptv} for the late-time data set displayed in Figs.~\ref{1_10},\ref{1_25} indeed show the characteristic of a shear-banded velocity profile similar to those reported in other studies\cite{masselon2008nonlocal,nghe2008high,masselon2010influence} - a central plug with $\dot\gamma = 0$ is flanked with sharp shear layers with non-zero $\dot\gamma$. While making quantitative comparisons between rheological parameters extracted from rheometric flow-curve and pipe-flow is often tricky \cite{callaghan1996study,masselon2010influence}, and we refrain from doing that in this paper, we note that the `high shear rate' band in the profiles plotted in Fig.\ref{1per_ptv} has $\dot\gamma \sim$ 8$s^{-1}$ which is close to the $\dot{\gamma_u}$ for 1\% VES as seen in Fig.\ref{flowcurve}. Generally, the shape of the velocity profile is not a certain indicator of shear banding in the system.\cite{cheng2017distinguishing}We will experimentally investigate the nature of these shear layers in the next section. Another feature apparent in Figs.~\ref{1_10},\ref{1_25}\ref{1per_ptv} is that the velocity profiles are not symmetric about the pipe centre in the plane analyzed. We don't have an explanation for this interesting feature, but asymmetric velocity profiles with a plug+shear-band were reported for the flow of entangled micellar systems under certain parameter ranges in a confined micro-channel.\cite{lutz2017situ} Those profiles corresponded to a stationary state, whereas our profiles are transient and might eventually develop into symmetric profiles. 
Additionally, these shear layers are fragile and appear and develop in a fairly intermittent and irregular manner. For the case of 1\% VES solutions which, by the end of the test, display a more well-behaved shear layer we see in Fig.~\ref{1_bet} that for intermediate times, the shear-layers for $Q$ = 0.1 ml/min (upper panel) are inverted when compared with that in Fig.\ref{1_10} with the layer of higher $\dot\gamma$ attached to the lower instead of upper wall whereas for $Q$ = 0.25  there are strong fluctuations in space and time as seen in Fig.~\ref{1_bet}'s lower panel. 

\subsection{Velocity Decay in Flow-Stoppage Tests }\label{decay}
In this last experimental results section, we report flow-stoppage tests where we stop the flow by abruptly setting flowrate to zero. Note, this stoppage flow is distinct from a flow in which forcing ($\Delta p$) is set to zero and resulting transient in decay in velocity can be then measured by solving the energy equation coupled to stress evolution.\cite{frigaard2019stability}Here, we set $Q=0$. In principle, this should ensure that the velocity becomes instantly vanishes everywhere (owing to incompressibility of the fluid) to satisfy continuity. However, in practice this doesn't happen as the flow runs through a semi-flexible connecting tube before flowing in the pipe. This tube deforms and relaxes according to changes in flow inside. This sets a delay in the system that accommodates finite flow that has can have a lifetime of a few seconds. In this section we show how this experimental artifact is valuable in highlighting flow-induced heterogeneity in VES solutions. More details of this test along with results for Carbopol and Glycerol are in Appendix C.

Provided the flow is homogenous (i.e no banded layers with difference in elastic properties), velocities in any region of the fluid should decay to zero at the same \textit{rate}, and that rate should be mimicked by the decay in flow rate as well as $Q \approx \bar U A$ where $\bar U$ and $A$ are the mean cross-sectional velocity and area respectively. We report decay experiments carried out with a PEO solution (which shows no shear-banding) as a control for the experiments carried out in this section.
\begin{figure}[htb!]
\centering
  \includegraphics[width=0.35\textwidth]{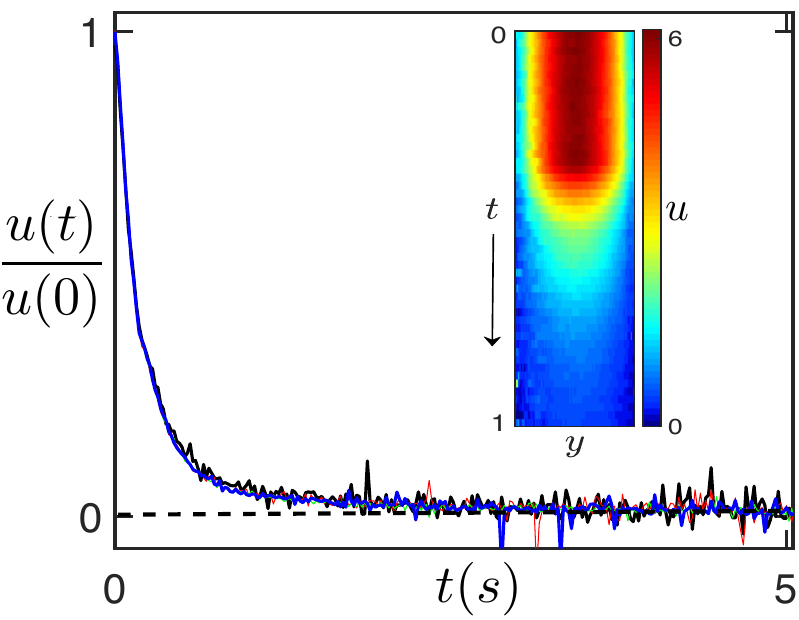}
  \caption{Decay of velocity for a flow stoppage test for PEO from $Q$ = 0.25 ml/min to 0 - rescaled velocity at $y$ = 0 (green), +0.3 (blue), -0.3 (red) and flowrate (black). Inset : time series of dimensional velocity showing complete decay process. }
  \label{peo_decay}
\end{figure}
Note, the PEO solution we used has a relaxation time of a few milli-seconds and is thus unlikely to affect the decay dynamics as a clear separation of time scales dictates that the delay mechanism that causes the finite flow post stoppage will dominate. We see, in Fig.\ref{peo_decay}, that for PEO, the decay rates are same for velocities measured in the centre or off-centre and these rates coincide with the decay in flow rate as well. In light of this, the behaviour displayed by 1\% VES in a flow stoppage test is remarkable. In Fig.\ref{edab_decay1} we plot the decay of flow rate and then velocity calculated by averaging the top,bottom 1/5 and middle 3/5 portion of the pipe cross section. It's evident that the middle section adjusts to the stopping of $Q$ very differently from the other sections and this is a clear manifestation of the VES's \textit{rheology}. 
\begin{figure}[htb!]
\centering
  \includegraphics[width=0.4\textwidth]{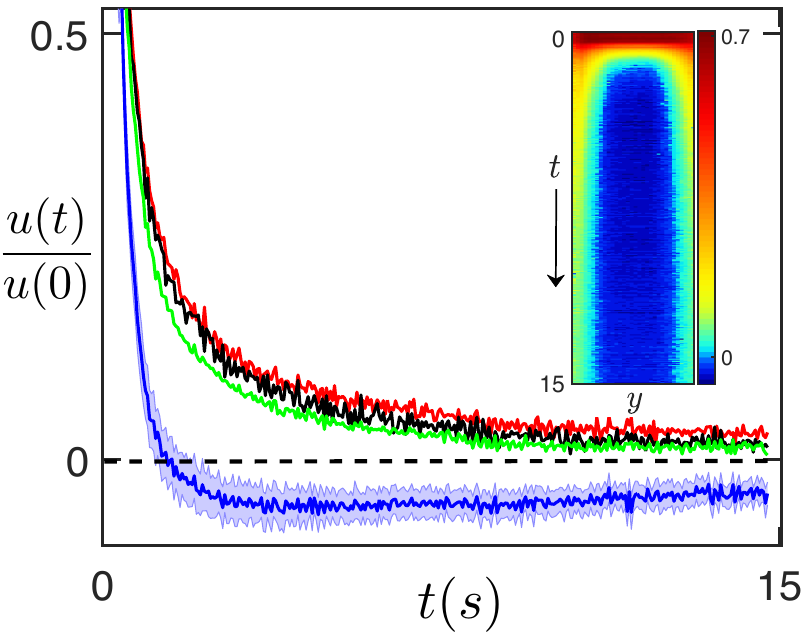}
  \caption{Rescaled velocity vs time for 1\% VES after $Q$ = 0.05 ml/min is set to 0 with different colours indicating decay of different quantities - top 1/5 (red), middle 3/5 (blue), bottom 1/5 (black) and flowrate (green). Shaded error bars are shown only for the blue curve to highlight that the negative velocities are numerically signficant. Inset : time series of dimensional velocity showing complete decay process.}
  \label{edab_decay1}
\end{figure}

\begin{figure}[htb!]
\centering
  \includegraphics[width=0.35\textwidth]{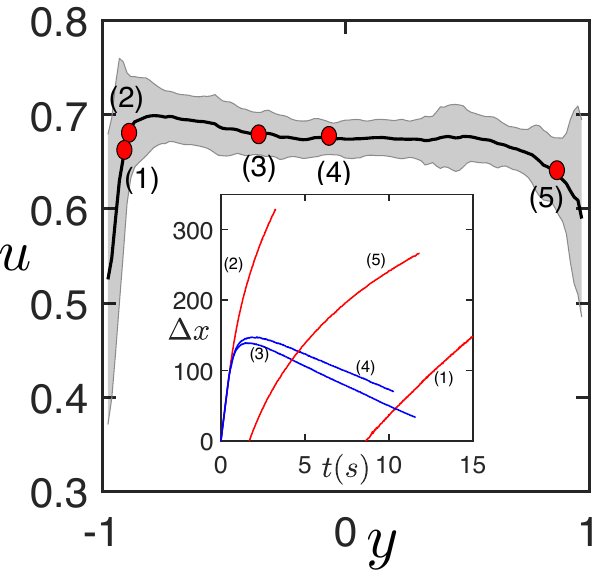}
  \caption{Time averaged velocity profile before initiating stoppage calculated by PIV for 1\% VES flowing at $Q$ = 0.05 ml/min. Inset : Trajectories of particles sourced from different regions with numbers indicating individual particles and red circles in main figure indicating the $y$ position in the pipe cross-section of the particle}
  \label{edab_decay2}
\end{figure}

We believe this process of differential decay is consistent with flow-induced heterogeneity in our system and suggests that the flow is made up of bands with different rheological properties that relax stress at different rates. Two features emerge from these dynamics: (1) The middle portion gains negative velocity before accelerating, (2) The velocity profile decay isn't monotonous and the initially concave shaped velocity near the wall (see profile in Fig.~\ref{edab_decay2}) becomes concave as observed in the velocity-front in the spacetime plot of the decay process in Fig.~\ref{edab_decay1}. %Both these features arise from the rheology of VES and bear similarity to the simulations of flow stopping (by setting forcing to 0) in a non-thixotropic fluid with finite elastic effects \cite{frigaard2019stability}. 
%The negative velocities attained by the middle layer are reminiscent of the elastic recoil effect reported in start-up flows in different complex fluids including wormlike micellar solutions. \cite{rassolov2020effects,rassolov2022role} The elastic recoil is in general a complex effect that is usually observed in start-up shear in a rheometric setting and can depend subtly on microstructural and rheological properties. \cite{rassolov2020effects,rassolov2022role} This deserves a systematic study which we don't undertake here. However, elasticity is a major driver of this effect and if these negative velocities are a result of an elastic recoil, it could occur because the central plug is only elastically deformed due to $\dot\gamma$ being 0 in the region. After stopping, it relaxes stress by snapping back like a band that is only elastically stretched, in contrast to the shear-bands which differ in viscoelastic properties. We note that such negative recoil post stoppage were observed with varying magnitudes in other flows of 1\% VES as well as some 2\% systems with shear-layers.  Well defined regions of positive and negative velocity are clearly seen in the vector plot in Fig.\ref{edab_decay_vector}. 

\begin{figure}[htb!]
\centering
  \includegraphics[height=6cm]{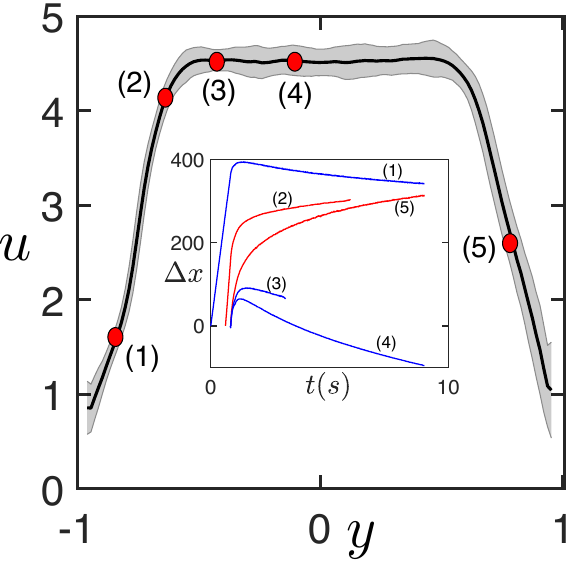}
  \caption{Time averaged velocity profile before initiating stoppage calculated by PIV for 1\% VES flowing at $Q$ = 0.25 ml/min after stepping down from $Q$ = 3 ml/min. Inset : Trajectories of particles sourced from different regions with numbers indicate individual particles and red circles in main figure indicating the $y$ position in the pipe cross-section of the particle. }
  \label{edab_decay3}
\end{figure}

Another way to visualise this differential decay process is by tracing the post-stoppage trajectory of particles picked from different regions of the fluid ($y$ coordinate). Basically we measure how their displacement (from their initial position) in the flow direction ($\Delta x$), changes in time. In Fig.\ref{edab_decay2} we see that the time-averaged velocity profile before stopping has a shear layer. For particles in that shear layer, $\Delta x$ increases with time, whereas for particles in the plug region, $\Delta x$ is non-monotonic in time indicating the acquisition of negative velocities. 
We use this method to highlight the complex and interesting arrangement of bands in another 1\% VES flow. We earlier demonstrated that for 1\% VES, velocity profiles can take asymmetric shapes and have a structure that could indicate multiple bands. An example of such a velocity profile is seen in Fig.\ref{edab_decay3} where the profile is asymmetric and the shear band attached to the bottom wall seems to be split into at least two distinct portions with different $\dot\gamma$. 

\begin{figure}[htb!]
\centering
  \includegraphics[width=0.4\textwidth]{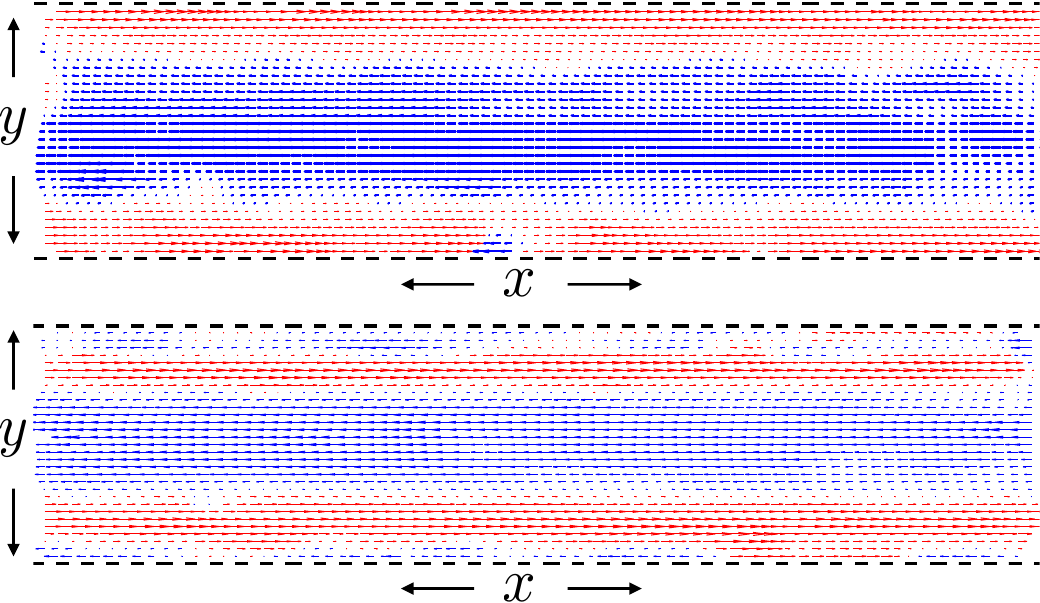}
  \caption{: A typical snapshot showing x-component of velocity during the relaxation post-stoppage for \% VES flowing at $Q$ = 0.05 ml/min (Upper) and $Q$ = 0.25 ml/min after step-down from $Q$ = 3 ml/min. Note, red arrows have positive velocities and blue arrows have negative velocities. Length of arrows indicate magnitude of velocity but aren't clearly indistinguishable.}
  \label{edab_decay_vector}
\end{figure}

When such a profile relaxes post stopping, there is a formation of alternating regions of plus/minus velocities as seen in the lower panel of Fig.\ref{edab_decay_vector}. Trajectories of particles from the shear-layer closer to the bottom wall can show both monotonic increase and non-monotonic variation of $\Delta x$ depending on which `band' they are sourced from, whereas those in the plug show non-monotonic variation. A particle from the shear layer close to the top wall shows increasing $\Delta x$ but few particles very close to the wall also show negative velocities as confirmed in the associated vector plot in Fig.\ref{edab_decay_vector} and visually confirmed in images of the flow.  We believe that this is a signature of transient banding with \textit{multiple} shear bands. Overall, our flow-stoppage tests provide evidence that the shear layers formed in the flow of wormlike micellar gels are shear bands that indicate a flow-induced heterogeneity. 

%Multiple bands for wormlike micellar solutions have been predicted \cite{zhou2012multiple} and reported before \cite{becu2004spatiotemporal,casanellas2015spatiotemporal,rassolov2022role}, but we believe ours is the first study to report and indirectly visualize this phenomenon in a pipe flow.  We admit that our flow-stoppage tests by setting $Q$ aren't ideal to rigorously probe these interesting banding scenarios and associated relaxation effects, but we agree with the assertion that properly constructed flow-stoppage tests can indeed by a very useful addendum to rheometric flows to understand these dynamical phenomenon and unentangle the non-trivial coupling of flow and flow-induced structural changes, as well as come-up with models to predict these behaviour.\cite{frigaard2019stability}
\section{Discussion}
We have shown that even flow in one of the more simple canonical geometries can show non-trivial features owing to the underlying complexity of the materials rheology.\cite{ewoldt2022designing} At first glance, the features of the flow dynamics of our WLM gels present as that of a yield stress fluid: at lower flowrates the fluid is `un-yielded' and hence flows as a plug, whereas at higher flow-rates/lower concentrations, the flow is made up of an unyielded plug with a yielded shear-layer region. However, we have shown that these flows must not be interpreted as those of a yield-stress fluid (such as Carbopol). First, the persistent plug flow uncovered in our studies is only a long-lived transient and part of the developing flow of a viscoelastic fluid. If we had access to much longer pipe lengths, these full-plugs we observe should give way to a finite shear-layer dictated by dynamics discussed in  Section.~\ref{3percent}. Indeed, channel flow for a shear-banding WLM solution has been shown to exhibit profiles with plugs and very thin layers where shear is concentrated, in experiments\cite{ober2011spatially} as well as in analytical calculations.\cite{castillo2020bulk} In the former, an augmented BMP model\cite{manero2007thermodynamic} is shown to display this velocity profile for fluids where structure relaxes fast and destruction is slow. These conditions seem to be satisfied for our WLM gel; the former is seen in the recovery profiles in Appendix B and the latter can be ascertained by noting that slowness of material destruction should be proportional to disentanglement time, which is expected to be large in our system. 

In the context of a developing viscoelastic flow, the pressure drop $\Delta p$ has a large contribution from the developing region. The value of $\Delta p$ corresponding to a steady state velocity profile, with lower wall shear stress and an increased thickness of the lubricating layer, should be much lower. There is also a question of value of the viscosity in this lubricating layer, the lower bound being the viscosity of water, $\eta_{w}$. It has been shown for other WLM systems, that this layer can have $\eta\gg\eta_{w}$, again increasing the thickness of the lubricating layer.\cite{nghe2010interfacially,kim2016transient} This \textit{low viscosity} phase coexisting with the plug can be caused by micellar migration from the near-wall to the bulk regions.\cite{kim2016transient}This process along with the growth in thickness of the lubricating layer to its eventual value can lead to said flow development in time with significant dissipative contributions to the measured $\Delta p$. In general, shear bands take a long time to develop, especially in light of long relaxation times for WLM gels.\cite{gupta2021rheology}. Whether the shear localization in the eventual shear layer is owing to a desorption or micellar disentanglement is an interesting question that we can't resolve in the current study. 

A small note on instabilities: all reported dynamics are from measured quantities in a single plane. As such we cannot systematically rule out the possibility of 3D flows that could be triggered by an elastic instability, \cite{larson1990purely} which although more likely in a set-up with curved streamlines (Couette), are not altogether impossible in a pipe flow like ours.\cite{castillo2020bulk} Discerning these effects is challenging owing to the difficulties in measuring 3D velocities in geometries but we note that we haven't observed a trend of particles going in and out of the plane (which would manifest as particles randomly appearing and disappearing in image snapshots) and can hence rule out instabilities in the bulk. Systems with high elasticity shear banding can also show instabilities owing to the region between the plug and shear layer getting destabilized.\cite{lerouge2006interface,castillo2020bulk} Experiments in a Couette cell have shown that wormlike micelles under shear display fluctuations over multiple time scales \cite{becu2004spatiotemporal,becu2007evidence} consistent with a growth and nucleation scenario for shear bands together with a non-trivial coupling between wall-slip and interface \cite{lettinga2009competition}. These dynamics manifest as fluctuations in $\dot\gamma$ at near the position of the shear-band interface (See for example Figs 2,4 in \citeauthor{becu2004spatiotemporal} and Figs. 10,11,16 in \citeauthor{becu2007evidence}). Over the period of a test, we notice large unsteadiness in wall-slip values (easily concluded from almost all time series presented that have a shear layer), and its likely that this coupling plays a role in the interface fluctuations reported by us as well. %undulations at the interface of plug and band

Here, we must also comment on background fluctuations in color visible in velocity time series (for example in Fig.~\ref{1_bet}). We have checked representative time series for Glycerol, PEO and VES - for the centreline velocity, for a fixed time period of recording, if we track the variation of $(u-\langle u \rangle/)\langle u \rangle$ (where $\langle u \rangle$ is a mean in time), the amplitude of the variation remains capped by $\sim$ 5\%. For Carbopol this variation is almost an order of magnitude lower, presumably due to its high viscosity and negligible elasticity. These background variations also have a dominant frequency which we think is not an artifact of data procesing but a physical oscillation that likely arises from the experimental apparatus. The dominant frequency in the time series for VES is smaller than those obtained for Carbopol and Glycerol solutions. This could be due to a nontrivial coupling between fluid elasticity and elasticity in the system (due to connecting tubes etc.) or the frequency of stepper motor working in the syringe pump.

Finally, we discuss the interesting features we observed in stopping the flow of VES, primarily the differential decay in velocities.  In particular, the negative velocities/flow-reversal examples we have reported in Sec.~\ref{decay} bring to mind the elastic recoil effect reported in start-up flows in different complex fluids including wormlike micellar solutions. \cite{rassolov2020effects,rassolov2022role} The elastic recoil is in general a complex effect that is usually observed in start-up shear in a rheometric setting and can depend subtly on microstructural and rheological properties.\cite{rassolov2020effects,rassolov2022role} It manifests as flow reversal when a drop in stress occurs post an overshoot in a shear startup test. Recently, a different WLM gel was shown to exhibit drastic negative velocities owing to elastic recoil, leading to the conclusion that elasticity and heterogeneity are necessary conditions for this effect.\cite{mccauley2023evolution} Our WLM gels are marked by high elasticity and the clear differences in how regions relax indicates a degree of heterogeneity in our system. Even though $\dot\gamma = 0$ in the plug, it does undergo elastic deformation. This is evidenced by noticing that the trajectories of particles taken from the plug region don't overlap (See insets of Figs.~(\ref{edab_decay2},\ref{edab_decay3}), clearly indicating that the plug has sustained deformation.  After stopping, it relaxes stress by recoiling back like a band in order to recover from its elastic deformation, in contrast to the shear bands which differ in viscoelastic properties and hence relax stress differently. This picture also explains why we didn't see this differential velocity decay effect in Carbopol where the amount of elastic deformation sustained in the plug is likely to be very small. We also note that a differential decay post stopping was observed with varying magnitudes in other flows of 1\% VES as well as some 2\% systems with shear-layers. But these weren't easily quantifiable and it is likely that the complex flow history during the pipe flow could have influenced the heterogeneity and hence the decay process.\cite{mccauley2023evolution}

%\textit{Is the strong wall-slip in 3\% and 2\% ($Q = $ 0.05,0.1) solutions a slip due to a layer of lubricating fluid or a shear-band?} We have argued that the plug profiles have a development length and (more importantly) pressures associated with it. This could be caused by development of shear-bands in time, the timescales for which could be very long. We need to revisit the values of $\Delta P$ and hence $\sigma_{w}$ used for the calculations for the wall-slip layer thickness $\delta$ earlier in Sections.~\ref{3percent} and \ref{2percent}. To find the $\delta$ for a full-developed plug-flow we would need to subtract out the dissipative contributions of the the development in $\Delta P$, leading to increase in values of $\delta$.. Typically, it is difficult to distinguish between the presence of a microscopic shear-band and depletion layer, both being methods of shear localization in thin regions \cite{mendez2003particle,ober2011spatially} and certainly we aren't in a position to do so. 

 \section{Conclusions}
In this paper we have studied the flow dynamics of a WLM gel in a canonical flow scenario using optical coherence tomography-based velocimetry. We showed that the velocity profiles in the observation window depend on the imposed flow rate and the concentration of the gel used and we observe stable and persistent plugs as well as plugs flanked by shear layers. However, crucially, these profiles also depend on time (and hence space) and must be interpreted as a developing viscoelastic flow. When full plugs transition to plugs+shear-layers they show undulations at the interface in between. We also observed asymmetric velocity profiles with multiple shear bands, perhaps the first time such profiles have been reported in pipe flows. Finally, we showed that the shear layers are manifestations of flow-induced heterogeneity by analyzing how velocities at different locations in the pipe decay after we stop the flow. 

The core message of our experiments is that the high elasticity of WLM gels can have a drastic effect on the flow. Ultra-long relaxation times of WLM gels\cite{gupta2021rheology} ensure that stress relaxation occurs on very long time (and hence length) scales. Hence, their flow can look like that of an `unyielded' plug flow of a yield stress fluid over length (and time) scales it takes to achieve steady state. This fact can be used to predict and engineer desired flows in different geometries. For instance, plug flows with thin shear layers have already been used as a method to prevent damage to cells.\cite{olsen2010yielding,blaeser2016controlling}. Controlling the size of the plug (and hence the shear layer) and having knowledge of the dynamics en route to the development of shear-layers can be beneficial in understanding these applications. Our experiments also highlight the signatures of high elasticity on flow dynamics, namely the highly irregular and unsteady nature of shear layer formation as well as flow-induced heterogeneity. Our results may benefit efforts to model the flow of complex materials that display such rheological signatures.\cite{de2012critical} Finally, there is value in studying the stopping flow of complex fluids as we have demonstrated in our case. It can be looked at as a flow test that can provide insight into how elasticity, for instance, can really alter flow profiles and behaviour.\cite{frigaard2019stability} It would be interesting to search for analogues of such a test with similar materials in other flows, for example a Couette flow in the rheometer.\cite{mccauley2023evolution} Our experiment reinforces WLM gels as an interesting complex material, distinct from other conventional WLM solutions. Indeed, we have cause to believe that owing to its high elasticity, it can tend to behave as soft glassy matter in certain flow scenarios.\cite{gupta2021rheology,mccauley2023evolution} Further work is needed if we want to understand the origins and effects of this particular behaviour. Such studies are expected to aid the growing research on yielding and flow of soft glassy materials.\cite{coussot2002coexistence,moorcroft2011age,divoux2013rheological,mccauley2023evolution} 

Regarding pipe flow there is ample scope for extension. For instance, experiments in longer pipes can allow us to access more developed profiles. Studying the dynamics of WLM gels in other geometries such as Hele-shaw cells, confined microchannels and even narrower round capillaries would allow us to access different regions of the flow curves. A controlled shear stress equivalent for a pipe flow can be attained by imposing pressure drop across the capillary instead of a flow rate. This is a natural extension of the experiments we reported here. Finally, its worthwhile to study the flow dynamics of wormlike micellar gels in any simple scenario because the rheological complexity guarantees rich dynamics. As WLM gels find use as tuneable soft materials in different contexts, there is a growing need to better characterize these dynamics. Some of this work has been already undertaken and we aim to report it in the future.\cite{gupta2023rheology}

%In all, our experiments have shed light on the transient fluid dynamics of wormlike micellar gels in simple geometries and will inform studies in which rheology and fluid dynamics of complex fluids are coupled in non-trivial ways. %WLM gels used for interesting applications. Many of these applications might involve pipe flows or flows in similar geometries. Out studies might be useful in design etc.....Also speak about Ian's code and possible results from it. 

\section*{Acknowledgements}
This research was made possible by research funding from Schlumberger and NSERC under the CRD program, project 505549-16. Experimental infrastructure was funded by the Canada Foundation for Innovation and the BC Knowledge Fund, grant number CFI JELF 36069. This funding is gratefully acknowledged. We thank Schlumberger Ltd. for providing us with the surfactant used for the experiments in the paper. R.G would like to thank Dr. Boris Stoeber and Dr. Mark Martinez as well as research students from their groups for facilitating access to labs where most of these experiments were conceived and carried out. R.G and M.D would also like to thank the following undergraduate students who had stints in the Complex Fluids lab at UBC and helped in the development of this project in various stages - Rohan Birk, Nile Waldal, Anastasia Vogl, Ivan Bao. R.G would also like to thank Cyprien Gay and Arjun Sharma for helpful discussions.

\section*{Appendix A : Benchmark Tests}\label{benchmark}
We show here the proper working of our experimental set-up and the robustness of methods and techniques employed. To demonstrate the capability of the pressure sensor integration with the capillary flow system we run a simple flowrate step-down test with pure glycerol solution in which we successively decrease the imposed flowrate $Q$ in a step-wise manner, pausing at each step. For Newtonian fluids, the Hagen–Poiseuille equation \citep{hplaw} predicts $\Delta p = \frac{8\eta L Q}{\pi {R}^4}$, where $\eta$ is the constant shear viscosity of the liquid and $R = D/2$ is the radius of the circular cross-section.  If we rescale $Q$ to absorb all the geometric dependencies in one parameter, $\hat Q$, we are left with a simpler equation $\Delta p = \eta \hat{Q}$ where $\hat Q = \frac{8QL}{\pi R^4} $. 
\begin{figure}[htb!]
\centering
  \includegraphics[height=5cm]{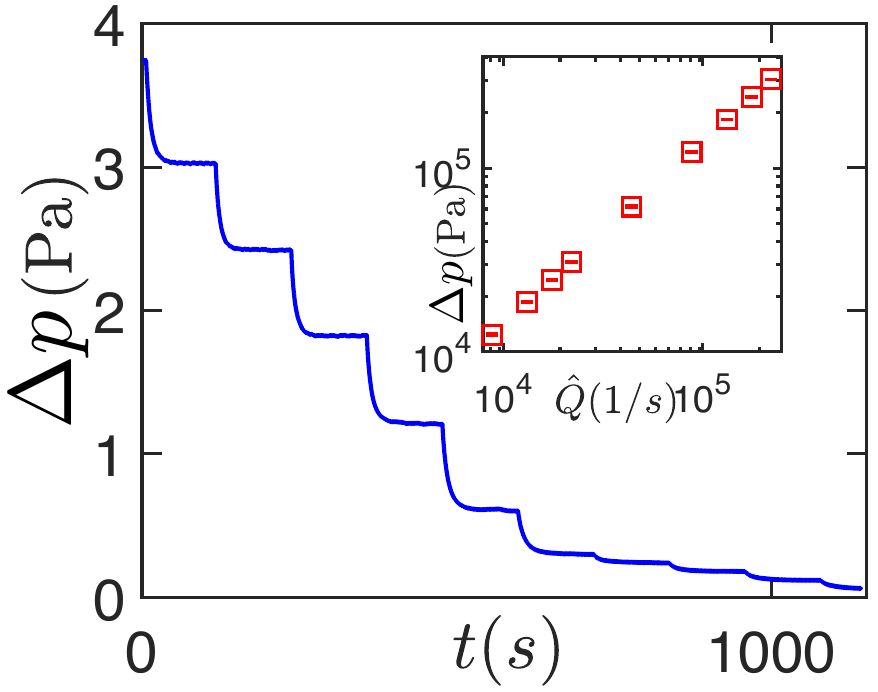}
  \caption{$\Delta p$ vs $t$ for pure glycerol in a $Q$ ramp down test at 20-21$^\circ$. Inset shows steady state $\Delta p$ extracted from main figure and plotted versus $\hat Q$. Note that x-axis in the main figure is in units of $10^5$ Pa.}
  \label{fig:glycerolp}
\end{figure}
In Fig.~\ref{fig:glycerolp} we plot the time series of $\Delta p$ for a step-down experiment in which we ramp-down in flowrate. We extract average values for $\Delta p$ for each step by taking the mean over the time-period in the step for which values of pressure drop remain constant and plot it vs $\hat Q$ in the inset of Fig.~\ref{fig:glycerolp}. The slope of the line joining the points in the inset gives us $\eta_{exp}$ = 1.39 Pa s$^{-1}$ that compares favourably with the actual $\eta$ for pure glycerol at 20$^\circ$ which is $\sim$ 1.41 Pa s$^{-1}$.\cite{glycerolpaper} In Fig.~\ref{glycerolv}, we plot the mean velocity profiles for the flow of a density matched solution of glycerol for two different flow rates. For Newtonian liquids, the shape of the velocity profile is independent of $Q$, and the velocity ($u$) scaled by the maximum centreline velocity ($u_{max}$) follows a parabolic variation w.r.t radial distance from centre, $r$ as - $\hat u = \frac{u}{u_{max}} = 1-{(\frac{2r}{D})}^2$. From Fig.~\ref{glycerolv} it is evident that the scaled velocity profiles ($\hat u$) for both $Q$ faithfully follow the expected Newtonian profile. 
\begin{figure}[htb!]
\centering
  \includegraphics[width=0.32\textwidth]{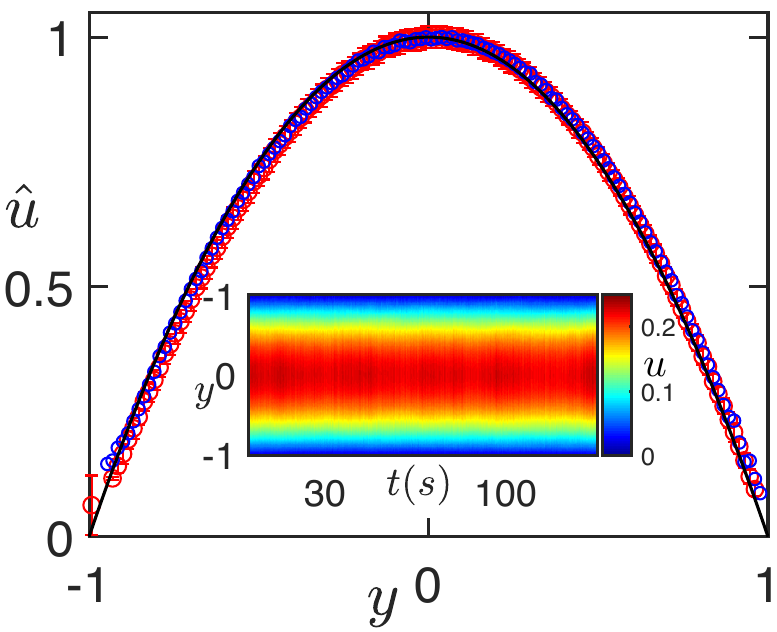}
  \caption{PTV results for $\hat u$ vs $y$ for the flow of density matched Glycerol solution at $Q$ =Ch4 0.01 (red) ,0.1 ml/min (blue). Note, error bars are not plotted for the latter. Solid black line is a parabolic profile. Inset : typical time series of velocity profile for $Q$ = 0.1 ml/min computed using spatially averaged PIV.}
  \label{glycerolv}
\end{figure}
\begin{figure}[htb!]
\centering
  \includegraphics[width=0.32\textwidth]{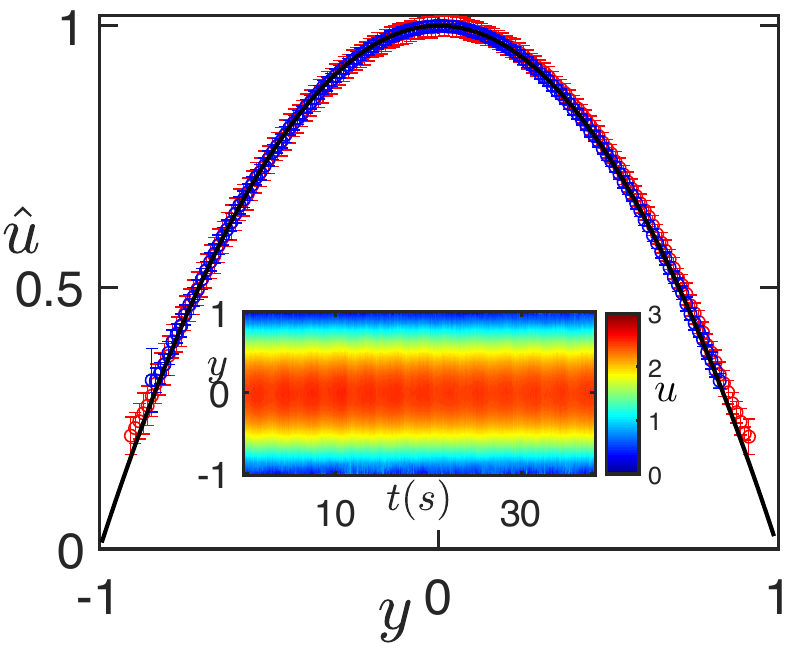}
  \caption{PTV results for $\hat u$ vs $y$ for the flow of 1\% PEO solution  at $Q$ = 0.1 (red) ,0.5 ml/min (blue). Solid black line is a parabolic profile. Inset : typical time series of velocity profile for $Q$ = 0.1 ml/min computed using spatially averaged PIV. }
  \label{peo}
\end{figure}

In Fig.~\ref{peo} we plot the velocity profiles for two flowrates $Q = 0,1,0.5$ ml/min for our PEO solutions. For polymeric solutions which are elastic and weakly shear-thinning and flows far from a transitional value of $Re$, the velocity profiles are also parabolic.\cite{peo1,peo2} The flows considered in this study are all for $Re$ much lesser than the transitional values reported in,\cite{peo2}and the PEO solution is only weakly shear-thinning in range of mean shear rates probed in this benchmark test, so we expect to see parabolic velocity profiles like the profiles experimentally obtained and plotted in Fig.~\ref{peo}. 
\begin{figure}[htb!]
\centering
  \includegraphics[width=0.45\textwidth]{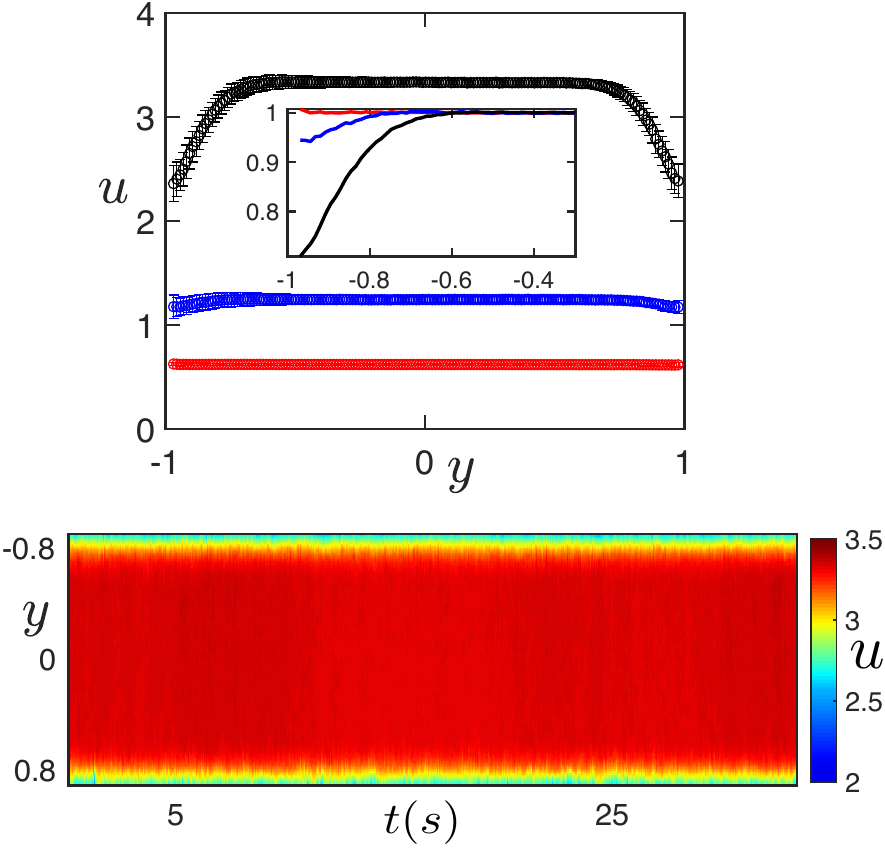}
\caption{Upper - PTV results for $u$ vs $y$ for $Q$ = 0.05 (red), 0.1 (blue), 0.25 (black) for Carbopol solution. Inset : $\hat u$ vs $y$ shown for $\sim$ half the pipe. Lower - typical time series of velocity profile for $Q$ = 0.25 ml/min computed using spatially averaged PIV.}
\label{carbopol}
\end{figure}
Finally, we show the velocity profiles for a 0.06\% Carbopol solution for three different values of $Q$, in Fig.~\ref{carbopol}. Carbopol is a microgel that is widely used as a model system for a yield-stress fluid.\cite{carbopol1} The characteristic features of the flow of such materials include (1) : A transition from a wall-wall plug to a plug with well-defined shear-layer with increasing $Q$ (or $\Delta p$), (2) : Decreasing size of the plug with increasing $Q$ (or $\Delta p$).\cite{carbopol1,carbopol2,masoudthesis} Both these features are reproduced in the velocity profiles shown in Fig.~\ref{carbopol}, with the decreasing size of plug with increasing $Q$ more evident in the inset in which we plot $\hat u$ for half the pipe. %As seen in Fig.~\ref{carbopol}, there is strong presence of wall-slip in the flow of these microgels, but as this section is intended as a demonstration of the proper working of our experimental set-up, we don't get into the details of it and refer the reader to other work.\cite{masoudthesis,daneshi2019characterising} Note, for this work, we define the plug region as the region where $\hat u$ is constant and $\dot\gamma \approx$0.

We must also mention that we have checked that the profiles displayed in this section are fully developed in space and time. Keeping fluid parameters constant, we measured velocity profiles in the same location at two different times during the test. We also measured the late-time velocity profiles at two different locations (500 and 700 mm from the inlet). For both cases, we found no change in the velocity profiles.

\section*{Appendix B : Flow dynamics in Step-down tests}\label{stepdown}

We conduct some step-down tests in which we change frowrate from 3 to 0.05,0.25 ml/min and measure velocities in intervals (typically one early-time and one late-time). An analogue test in a rheometer is generally a step-down in $\dot\gamma$ though it's difficult to make a proper one-to-one comparison owing to the difference in geometries. For instance, in a pipe, the fluid contained experiences a range of shear rates and hence cataloguing a response and attributing to specificities like thixotropy is a challenging task. We have noted that our wormlike micellar gels show strong time-dependence and preliminary rheology step-down tests show signatures of thixotropy such as a buildup in shear-stress post drop-off in shear rate (although these effects were noted for more concentrated VES).\cite{gupta2021rheology}We expect that in the pipe, we should notice a build-up in structure after the micellar structure is presumably broken down in large portions of the pipe due to a high value of imposed flowrate. However, post-structure recovery, the fluid elements should evolve in-time as they would in a pipe flow at the stepped down $Q$, albeit with a pre-imposed strain owing to the \textit{pre-shear} of a high $Q$. 
\begin{figure}[htb!]
\centering
  \includegraphics[width=0.47\textwidth]{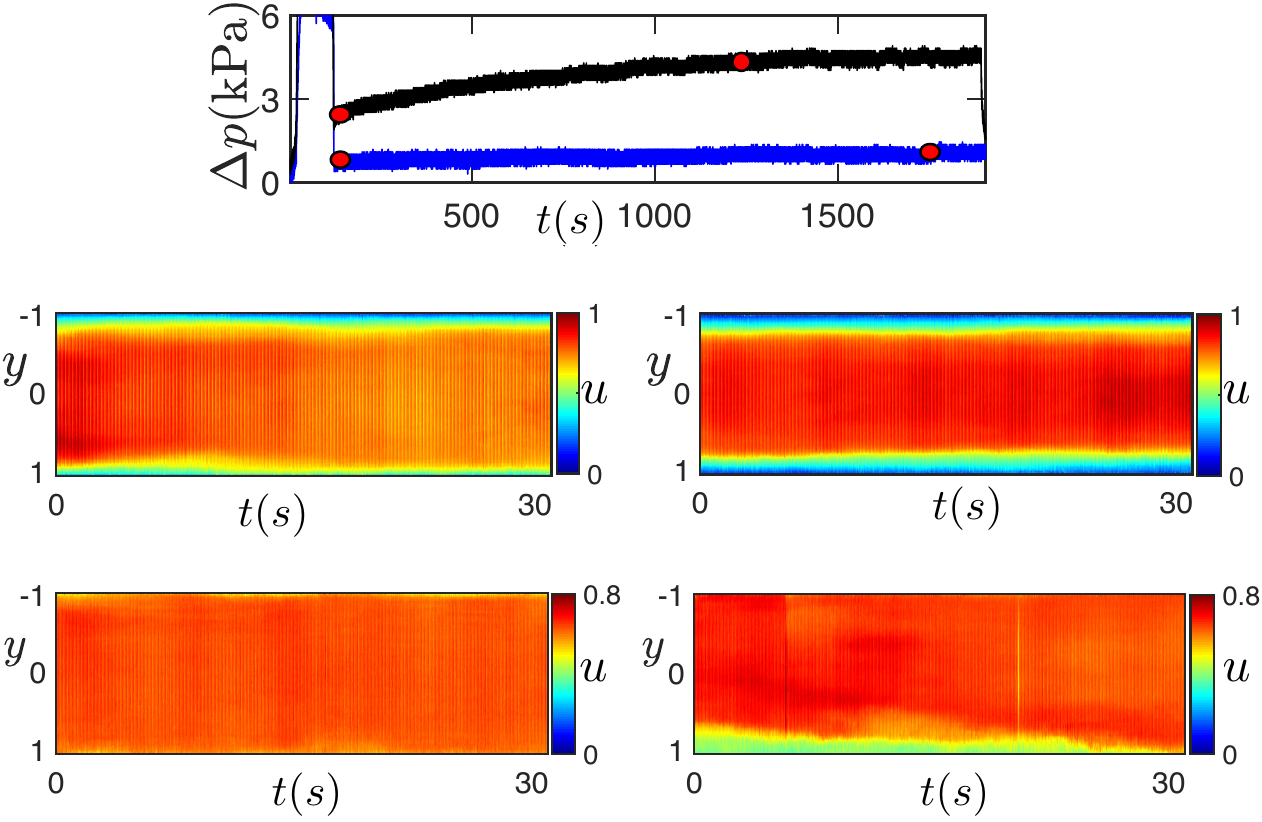}
  \caption{Dynamics of step-down from $Q$ = 3 to 0.05 ml.min for 1\% and 2\% VES. Top : $\Delta p$ vs $t$ for 1 (blue), 2\% VES (black). Middle : velocity time series for early (left) and late time (right) for 1\%. Bottom : velocity time series for early (left) and late time (right) for 2\%. Red circles in the top panel indicate the time from which time series are sourced.}
\label{step_005}
\end{figure}
\noindent
This should, in principle, allow us to access more developed profiles than in tests where we impose a fixed flowrate. This picture is however complicated by typical build-up dynamics and timescales involved. As is now routine in modelling fluids with time dependent structure, a construction/destruction parameter is used which is usually shear rate dependent.\cite{de2012critical,ewoldt2017mapping}Transient velocity profile evolution should be dictated by the relative magnitude of these terms, which in term should atleast depend on the combination of concentration and flowrate employed in a flow test. 
\begin{figure}[htb!]
\centering
  \includegraphics[width=0.47\textwidth]{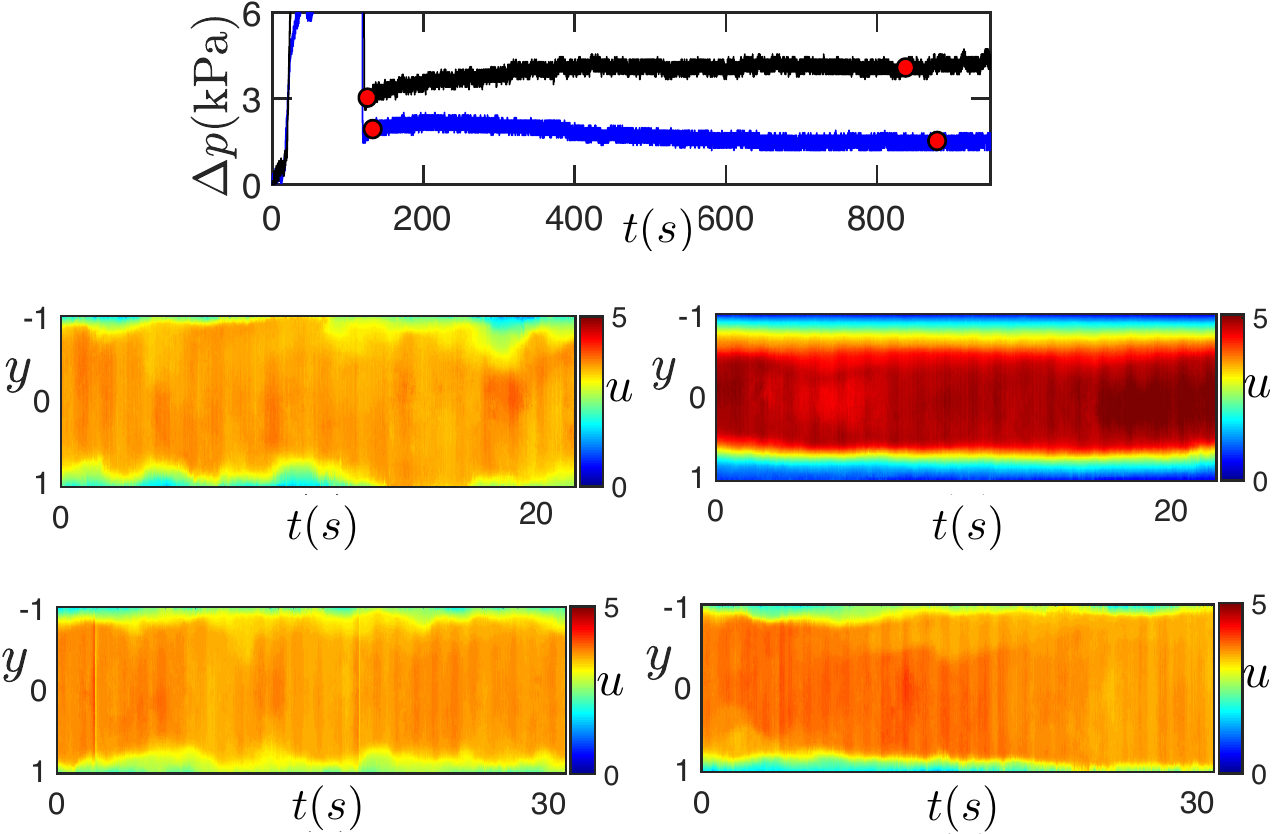}
\caption{Dynamics of step-down from $Q$ = 3 to 0.25 ml.min for 1\% and 2\% VES. Top : $\Delta p$ vs $t$ for 1 (blue), 2\% VES (black). Middle : velocity time series for early (left) and late time (right) for 1\%. Bottom : velocity time series for early (left) and late time (right) for 2\%. Note the red circles in the top panel indicate the time from which time series are sourced.}
\label{step_025}
\end{figure}
For example, for a step down to $Q$ = 0.05 ml/min, a few seconds (5-7 s) after step-down, while the 1\% VES solution (Fig.~\ref{step_005}, middle-left) recovers to profile with a distinct plug and shear-layer, the 2\% solution (Fig.~\ref{step_005}, bottom-left) recovers to a plug with a negligibly thin shear layer. At this point, we must mention that the pre-shear at $Q$ = 3 ml/min may not act as an equal pre-shearing condition for both, but we believe that the difference in post-recovery velocity profiles is majorly guided by the differences in $c$ here. Note that at this low $Q$, we had earlier shown that 2\% VES tends to form stable persistent plugs. But we see here that the change in initial conditions has allowed the fluid to show the presence of a shear-layer at late time. This shear-layer is highly spatially and temporally irregular. The time series measured before the late time one shown Fig.\ref{step_005} (bottom-right) has little to no presence of a shear-layer (not shown), whereas for the time series following it, a shear-layer is clearly present. For the case of 1\% VES, the recovery is followed by a developing shear-layer in time, similar to dynamics in an imposed flowrate tests, but with a different timescale and velocity profile, as is expected from the change in initial shearing condition. 

In Fig.~\ref{step_025}, we see that for 1\% VES solutions (middle panel), again there is a typical evolution of a shear-layer with increasing time. It's also evident in Fig.~\ref{step_025} that the post-recovery profiles for both 1\% and 2\% VES solutions are similar, but in the former, a well-defined shear layer eventually develops (middle-right) while the latter, at late-time, still shows the presence of a fluctuating shear-layer (bottom-right) that hasn't changed markedly from the post-recovery profile. In contrast, 3\% solution quickly recovers to nearly a full-plug (similar to the 2\% early-time profile in Fig.\ref{step_005}) when stepped down to $Q$ = 0.25 ml/min, and the plug persists without a consistent presence of a shear-layer even at late-times. These results underscore the strong dependence of recovery and evolution dynamics of these flows on $c,Q$ employed.  

\section*{Appendix C : Flow-Stoppage Tests with Carbopol and Glycerol}
\begin{figure}[htb!]
\centering
  \includegraphics[width=0.35\textwidth]{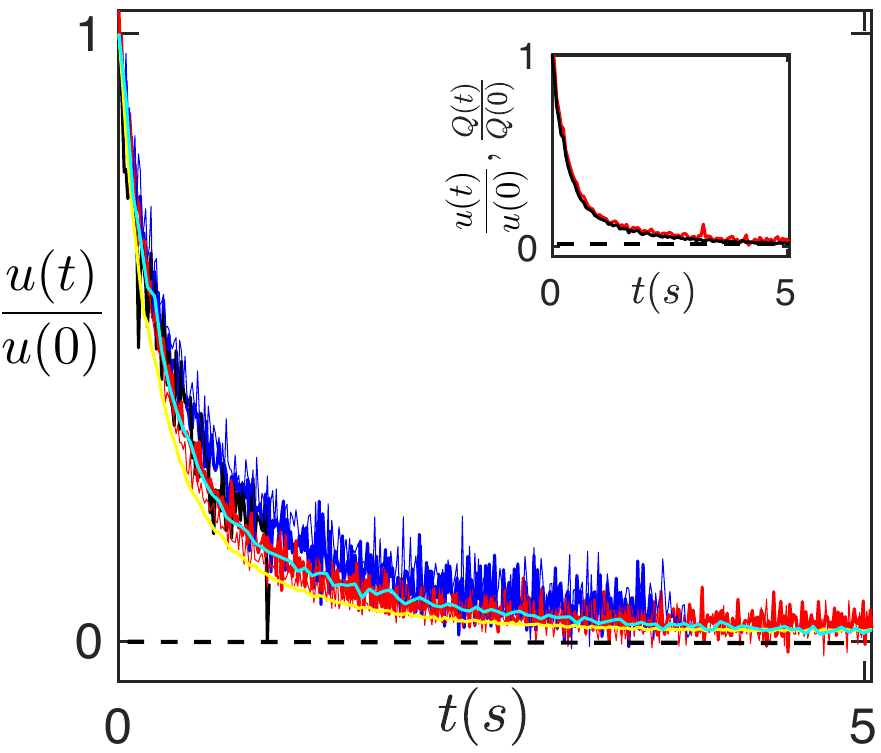}
  \caption{Collection of trajectories and averaged quantities that show decay of rescaled velocity with time - Centre-averaged Carbopol (yellow), centre-averaged Glycerol (cyan), individual particles in Carbopol : $y$= 0.3,0.4 (red), 0.62,0.65 (blue), 0.96 (black). See black in Fig.~\ref{carbopol} to spatial regions corresponding to these y. Inset : Decay plots for Glycerol - centre-averaged (red), flowrate (black). }
  \label{carb_decay}
\end{figure}

This test concerns fast transients and the resultant dynamics can be affected by factors other than rheology of the test fluid. For instance, there is a timescale associated with the syringe pump's stepper motor response and an visco-inertial timescale for the fluid which is a result of viscous damping. The former is of $O(10 \mu s)$ and for $Re$ in operational range used used, the latter has a relaxation timescale of $<$ 1s. In Fig.\ref{carb_decay}, we plot the decay of velocity in stopping tests for Carbopol (initial $Q$ = 0.25 ml/min) and Glycerol (initial $Q$ = 0.05 ml/min). For Carbopol we pick three types of trajectories, one corresponding to particles in the plug, particles near the transition layer, and particles in the shear-layer. Note, since the particle based $Re$ i.e Stokes number is very small, it effectively has no inertia and faithfully follows the local velocity of the fluid. We also run a coarse mesh PIV for the Carbopol and Glycerol image-sets and the velocity decay for these are obtained from averaging the velocity in a spatial region around the centre. 

In all the decay curves shown in Fig.\ref{carb_decay} the decay in velocity is in $O(s)$ (between 2-3 s) and curiously all the curves lie on top of each other (minor diffferences can be owing to trajectory selection, data noise and spatial averaging). The observation of $O(s)$ dynamics allows us to neglect viscous decay and stepper motor change induced delay and look for another larger timescale. For no differential decay after stopping the flow we claimed (in Sec.~\ref{decay}) that - (1) : rate of velocity decay should be the same everywhere in the fluid and (2) : This rate should be the same as the decay rate in flowrate. The former property is seen in the overlap of different carbopol trajectories in Fig.\ref{carb_decay} and the latter in the inset of Fig.\ref{carb_decay} where the curves for velocity and flow-rate decay for glycerol solution are the same.

%%%END OF MAIN TEXT%%%


\bibliography{biblio}% Produces the bibliography via BibTeX.

\end{document}